   \documentclass[12pt,preprint]{aastex}

   \usepackage{graphicx}
   \usepackage{ulem}

\newcommand{\kmps}{km s$^{-1}$}
\newcommand{\cc}{cm$^3$}
\newcommand{\gcc}{g cm$^{-3}$}
\newcommand{\Brute}{\texttt{Brute}}
\newcommand{\Synow}{\texttt{Synow}}
\newcommand{\Nifs}{$^{56}$Ni}
\newcommand{\Fefs}{$^{56}$Fe}
\newcommand{\gammaray}{$\gamma$-ray}

\newcommand{\lamang}{\lambda_{\textrm{\scriptsize \AA}}}
\newcommand{\caii}{\ion{Ca}{2}}
\newcommand{\wl}{$\lambda$}
\newcommand{\ww}{$\lambda\lambda$}
\newcommand{\handk}{H\&K}
\newcommand{\halpha}{H$\alpha$}
\newcommand{\hbeta}{H$\beta$}
\newcommand{\hgamma}{H$\gamma$}

\begin{document}

\title
{
 On the Geometry of the High-Velocity Ejecta of the Peculiar
 Type Ia Supernova 2000cx
}

\author
{
 R. C. Thomas\altaffilmark{1,2}, 
 David Branch\altaffilmark{1},
 E. Baron\altaffilmark{1},
 Ken'ichi Nomoto\altaffilmark{3},
 Weidong Li\altaffilmark{4},
 and Alexei V. Filippenko\altaffilmark{4}
}

\altaffiltext{1}
{
 Department of Physics and Astronomy,
 University of Oklahoma,
 Nielsen Hall RM 131,
 Norman, OK 73019
 (thomas,branch,baron)@mail.nhn.ou.edu
}

\altaffiltext{2}
{
 Present address: Lawrence Berkeley National Laboratory, 
 1 Cyclotron Road, MS 50R5008,
 Berkeley, CA 94720-8158
}

\altaffiltext{3}
{
 Department of Astronomy \& Research Center for the Early Universe,
 University of Tokyo, Bunkyo-ku, Tokyo, 113-0033, Japan
}

\altaffiltext{4}
{
 Department of Astronomy,
 601 Campbell Hall, 
 University of California, Berkeley, CA 94720-3411
}

\begin{abstract}
High-velocity features in Type Ia supernova spectra provide a way to 
probe the outer layers of these explosions.  The maximum-light spectra
of the unique Type Ia supernova 2000cx exhibit interesting \caii\ 
features with high-velocity components.  The infrared triplet absorption
is quadruply notched, while the \handk\ absorption is wide and flat.
Stimulated by a three-dimensional interpretation of similar \caii\
features in another Type Ia supernova \citep[SN 2001el,][]{Kasen2003}, we
present alternative spherically symmetric and three-dimensional ejecta
models to fit the high-velocity ($v > 16,000 $ \kmps) \caii\ features
of SN 2000cx.  We also present simple estimates of the high-velocity
ejecta mass for a few trial compositions and discuss their implications
for explosion modelling.
\end{abstract}



\section{Introduction}

One-dimensional (1D) white dwarf explosion models like W7
\citep{Nomoto1984} have been used to synthesize spectra in good
agreement with those observed of Type Ia supernovae 
\citep[SNe Ia; see][for a general review]{Filippenko1997}.  But
recent polarization observations of the SNe Ia
1997dt \citep{Leonard2000b},
1999by \citep{Howell2001},
2000cx \citep{Leonard2000a},
and 2001el \citep{Wang2003, Kasen2003} indicate that 
at least some SN Ia envelopes deviate significantly 
from spherical symmetry.

The strongest case for such deviation appears in the \caii\ lines of
SN 2001el.  This event exhibits an unusual \caii\ infrared (IR)
triplet in its flux spectrum about a week before maximum brightness.  
Two triplets are evident:
one corresponds to photospheric-velocity (PV) material and the other
to higher-velocity (HV) material.  In the polarization spectrum, the
HV feature coincides with significant intrinsic net polarization.
\citet{Kasen2003} investigate a number of envelope models to account
for the HV feature.  Generally, they conclude that (1) incomplete
covering of the photosphere and (2) some deviation from spherical
symmetry are required to produce the HV features.  Unfortunately,
conditions did not permit simultaneous observation of the blue
spectrum so a corresponding phenomenon in the \caii\ \handk\ feature
remains unconfirmed.

Another unusual \caii\ feature is found in the spectrum of the
peculiar SN 2000cx \citep{Li2001}.  Near maximum light, its ostensible
\caii\ IR triplet possesses an interesting quadruply notched feature
perhaps due to what Li et al.\ call ``some unique distribution of Ca
in the ejecta of SN 2000cx,'' extending to high velocity.  The
wavelength coverage of the near-maximum light spectra is excellent.
Though polarization data at the same wavelengths are not available, 
simultaneous fits of the \handk\ and IR triplet features may help
constrain at least the HV photospheric covering fraction (if indeed
the HV material is not distributed in spherical symmetry).

It remains unclear why clumpy HV ejecta could occur in SNe Ia.
Detailed synthetic spectropolarimetry has yet to test explosion models
free from the constraint of spherical symmetry \citep{Khokhlov2000,
Reinecke2002, Gamezo2003}.  These explosion calculations have 
neither proceeded
to the free-expansion phase nor provided abundance distributions
required for detailed spectrum synthesis.  On the other hand,
methods for calculating synthetic spectra from three-dimensional (3D)
models of SNe are in their infancy \citep{Thomas2002, Kasen2003}.
Hence parameterized, direct analysis of observed SN spectra remains a
powerful way of guiding explosion modellers to replicate geometrical
phenomena in their models.

In this article, the goal is to analyze the \caii\ features in the
spectrum of SN 2000cx at a single epoch to constrain the HV ejecta
geometry.  A separate multi-epoch, but exclusively 1D analysis of
this object is forthcoming (D. Branch et al., in preparation).  That
work addresses such orthogonal problems as the exhaustive identification
of PV features and the unusual color evolution of this SN.

The outline of the remainder of this article is as follows.  In \S2 we
present the spectra and identify the features of interest.  In \S3 we
use the 1D direct analysis code \Synow\ and the analogous 3D code
\Brute\ to test a few geometrical distributions of \caii\ line optical
depth.  We discuss the candidate models and estimate the HV ejecta
mass in \S4.  Conclusions appear in \S5.


\section{Spectra}

\begin{figure}
\centering
\includegraphics{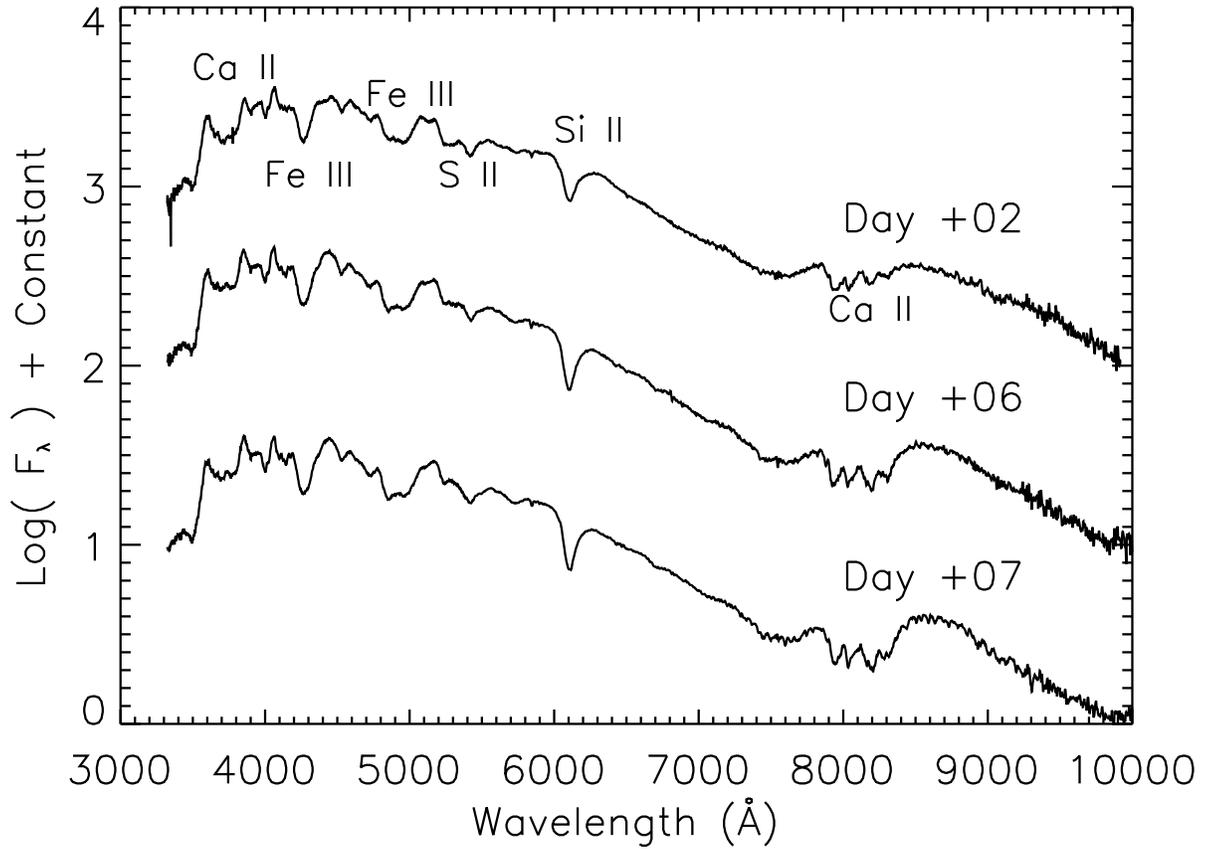}
\caption
{
 Near-maximum light spectra of SN 2000cx \citep{Li2001}.  The epoch 
 relative to maximum light is listed above each spectrum.  All spectra
 shown in this paper have been corrected for the redshift of the host
 galaxy, NGC 524, $z = 0.0080$.
 \label{fig:timeSeq}
}
\end{figure}

In Figure \ref{fig:timeSeq} are three spectra of SN 2000cx collected
near maximum light, originally presented by \citet{Li2001}.  The usual
SN Ia absorptions from \ion{Si}{2} (near 6150 \AA) and \ion{S}{2}
(near 5400 \AA) are accompanied by absorptions due to
\ion{Fe}{3} (near 4300 \AA\ and 5000 \AA).  There is very little, if
any, signature of \ion{Fe}{2}.

Between 7900 \AA\ and 8400 \AA\ is a series of four notches.  Usually
at this phase, only absorption from the two stronger lines of the
triplet (\ww 8542, 8662 --- often blended) are visible in this region,
and these are the best candidates for the two redder notches.  The two
bluer ones are perfectly consistent with a \caii\ IR triplet shifted to about
22,000 \kmps\ toward the observer.  The bluest and weakest line of the 
\caii\ IR triplet (\wl 8498) is approximately ten times weaker than 
\wl 8542, and is likely heavily blended with that line.
Other ions are unlikely to produce
the notches; candidates such as \ion{O}{1} are unconvincing due to
the absence or weakness of concomitant lines in the spectra.

The corresponding \caii\ \handk\ absorption feature (3500 \AA\ to 3800 \AA)
is wide and flat.  A collection of currently unidentified narrow absorption
features obliterates its emission peak.

\begin{figure}
\centering
\includegraphics[scale=0.85]{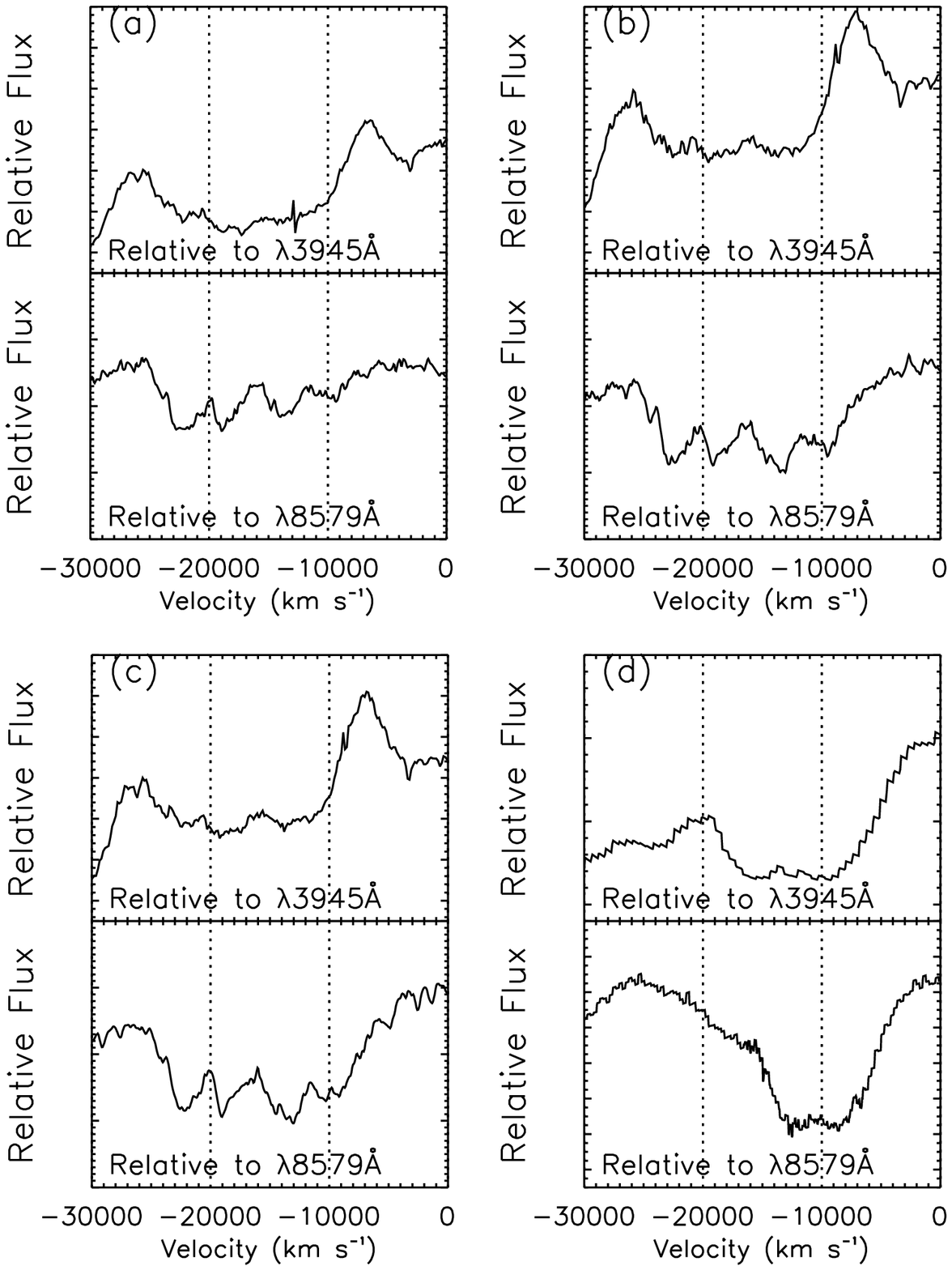}
\caption
{
 \caii\ features in SN 2000cx at days 2 (a), 6 (b), and 7 (c) after
 maximum light, and those in SN 1994D at day 3 (d) after maximum.
 The features are plotted in terms of velocity
 relative to the observer, using the $gf$-weighted wavelengths of
 the \caii\ \handk\ and IR triplet features.
 \label{fig:velPlot}
}
\end{figure}

Figure \ref{fig:velPlot} displays the \caii\ features of SNe 2000cx
and 1994D near maximum light in terms of velocity relative to the
observer.  The top half of each panel is the \handk\ feature relative
to the $gf$-weighted doublet wavelength (3945 \AA), while the bottom
half is the IR triplet relative to its $gf$-weighted wavelength (8579
\AA).

The fact that these two features are blends complicates the issue of
choosing a reference wavelength for producing Doppler-space plots.  In
a blend of P-Cygni lines, redder components dominate the shape of the
aggregate profile if they are strong; they screen out the bluer lines.
Nevertheless, the bluer lines must have some effect since they feed
radiation to the redder lines.  Given these ambiguities, we
compromise by using the $gf$-weighted wavelengths for reference.

Figures 2a, 2b, and 2c are from SN 2000cx at days 2, 6, and
7 past maximum light, respectively.  For comparison, the normal SN
Ia 1994D \caii\ features at day 3 past maximum appear in Figure
2d.  \citet{Hatano1999a} use HV \caii\ and \ion{Fe}{2} to improve
their synthetic fits of SN 1994D.  In that SN, the effect of the HV
\caii\ is much less pronounced than in SN 2000cx.  In fact, the IR
triplet of SN 2000cx has a HV absorption with a depth relative to
the continuum between that of SNe 1994D and 2001el.

Note that the four notches in the SN 2000cx IR triplet absorption do
not evolve appreciably in velocity space over time, and that the
overall velocity ranges of both features match.  There appears to be a
case for a one-to-one correspondence between notches in the IR triplet
and depressions in the \handk\ absorption, but the blended nature of
both features makes absolute confirmation difficult.  Furthermore,
the mild fluctuations visible in the \handk\ absorption could easily
be due to some weak, narrow lines superposed on the \caii\ feature.

We henceforth designate the parts of the \caii\ features forming above
16,000 \kmps\ relative to the $gf$-weighted wavelengths of \handk\ and
the IR triplet as the ``HV features.''  Our fitting strategy will be to
concentrate on replicating the velocity range of these features and
some of their structure.  Since the three SN 2000cx spectra in Figures
\ref{fig:timeSeq} and \ref{fig:velPlot} are all quite similar, we
restrict the focus of the remainder of this paper to the spectrum
obtained two days past maximum light.


\section{Fits}

The simplest conceptual model of a SN Ia between a few days to months
after explosion consists of an homologously expanding envelope
surrounding a continuum-emitting, optically thick core.  The decay of
freshly synthesized \Nifs\ to \Fefs\ at the center of the SN releases
energy in the form of \gammaray s.  This energy diffuses outward and
degrades to ultraviolet, visible, and infrared wavelengths via
interactions with the matter in the SN atmosphere.

This elementary model has been implemented in a variety of codes
intended for the empirical analysis of SN spectra \citep{Jeffery1990,
Fisher2000, Mazzali1993}.  These codes all implement a sharply defined
photosphere at some radius as a lower boundary for a line-forming
envelope.  They utilize the Sobolev approximation \citep{Sobolev1947,
Castor1970, Rybicki1978}
to construct radiation field estimates for given optical depth
distributions.  Optical depths may be parameterized (in velocity space
and wavelength) or determined from a selected abundance and density
model with or without a self-consistent temperature structure.

These codes have proved useful for fitting observed spectra to
constrain the structure of SN envelopes.  This empirical process is
called ``direct'' analysis to distinguish it from ``detailed''
analysis where the full radiative transfer and non-local thermodynamic
equilibrium (NLTE) rate equations are solved \citep{Hauschildt1999}.
Specifically, the purpose of direct analysis is (1) to determine what
species are present in a SN line-forming region and (2) to constrain
the velocity space distribution of those species.  The assumptions and
approximations used generally restrict direct analysis to line
phenomena, in particular to the Doppler shifts and overall shapes of
absorption features.  The results of direct analysis provide useful
guidance to detailed spectrum modellers and to explosion modellers as
well.

\Synow\ is a direct analysis code that has been used to fit many
spectra of various types of SNe, e.g., SN Ia 1994D
\citep{Hatano1999a}, several SNe Ib \citep{Branch2002}, SN Ic
1994I \citep{Millard1999}, and SN II 1999em \citep{Baron2000}.
\Synow\ relies on spherical symmetry and parameterized line optical
depth.  Each selected ion is assigned an optical depth in a
``reference'' line, usually a strong optical line.  All other line
optical depths for the same ion are determined assuming 
Boltzmann excitation
of the levels at some specified temperature.  The optical depths are
scaled as a function of radius to fall off exponentially or according
to a power law.  The full details of \Synow\ are described by
\citet{Fisher2000}.

We have recently developed a code similar to \Synow\ in its degree of
parameterization, but free from the constraint of spherical symmetry.
This code, \Brute, is based on Monte Carlo techniques presented in a
series of papers by Lucy and Mazzali \citep[e.g.,][]{Mazzali1993}, and
will be discussed in detail elsewhere (R. C. Thomas, in preparation).
In \Brute, spatial parameterization of line optical depth is managed
through a 3D template.  To establish the radiation field estimates in
all transitions throughout the envelope, Monte Carlo energy packets
are emitted from the core and their scattering histories are followed.
The required memory for \Brute\ is much larger than that for \Synow, 
and \Brute\ also lacks the speed of its 1D counterpart.

\subsection{\Synow\ Fits}

Here we investigate a few spherically symmetric distributions of
\caii\ optical depth.  The assumption of homology in the SN atmosphere
($v \propto r$) permits us to parameterize Sobolev optical depth in
terms of velocity $v$ relative to the explosion center.  In spherical 
symmetry, we may designate domains $v_{min} < v < v_{max}$ within which
we define reference line Sobolev optical depth according to the rule
\begin{equation}
 \begin{array}{r@{\quad \quad}l}
  \tau_{ref}(v) = \tau_{ref}( v_{min} ) 
                  \exp [ ( v_{min} - v ) / v_e ]
  &
   (v_{min} < v < v_{max}), \\
 \end{array}
 \label{eq:sphRule}
\end{equation}
where $v_e$ is an $e$-folding length.  If $v_{min}$ is greater than
the velocity at the photosphere $v_{ph}$, we say the optical depth
profile is {\it detached}.  Outside of the domain we may set the
reference optical depth to zero, or use another rule for the same 
reference line to set up a superposition of optical depth profiles.
Other line optical depths for the same ion are assigned assuming
excitation temperature $T_{exc}$.  The velocity at the photosphere in
all the presented \Synow\ fits is $v_{ph} = 12,500$ \kmps.

\Synow\ uses a blackbody-emitting photosphere, clearly insufficient to
account for all real continuum processes at work in the formation of 
a SN spectrum.  This limits the range over which the synthetic continuum 
level can be made consistent with that observed, making fits of entire spectra
extending from 3000 \AA\ to 10,000 \AA\ problematic.  We adopt a
piecewise approach and choose a convenient blackbody temperature
applicable for the blue part of the spectrum.  Fitting only the major
features blueward of 6000 \AA\ (excluding lines of \caii), we find
that a blackbody temperature $T_{bb} = 12,000$ K yields a decent fit to
this region.  Since increasing $T_{bb}$ from this value changes the
continuum slope in the neighborhood of the IR triplet only slightly, we
choose to merely scale the synthetic spectra down in order to fit
that feature.

Except for the optical depth parameters of \caii, the other parameters
($v_{ph}$, $T_{bb}$, and optical depths of ions listed in Table
\ref{tab:noCa}) stay fixed from fit to fit.  Since the \caii\ IR
triplet may blend with some lines of \ion{O}{1}, a small amount of
optical depth for that ion is included in the fit, but its presence
has little impact on the results.  The parameters used for the
various \caii\ optical depths are listed in Table \ref{tab:synow}.

\subsubsection{One-Component Fits}

\begin{figure}
\figurenum{3}
\epsscale{1.00}
\plotone{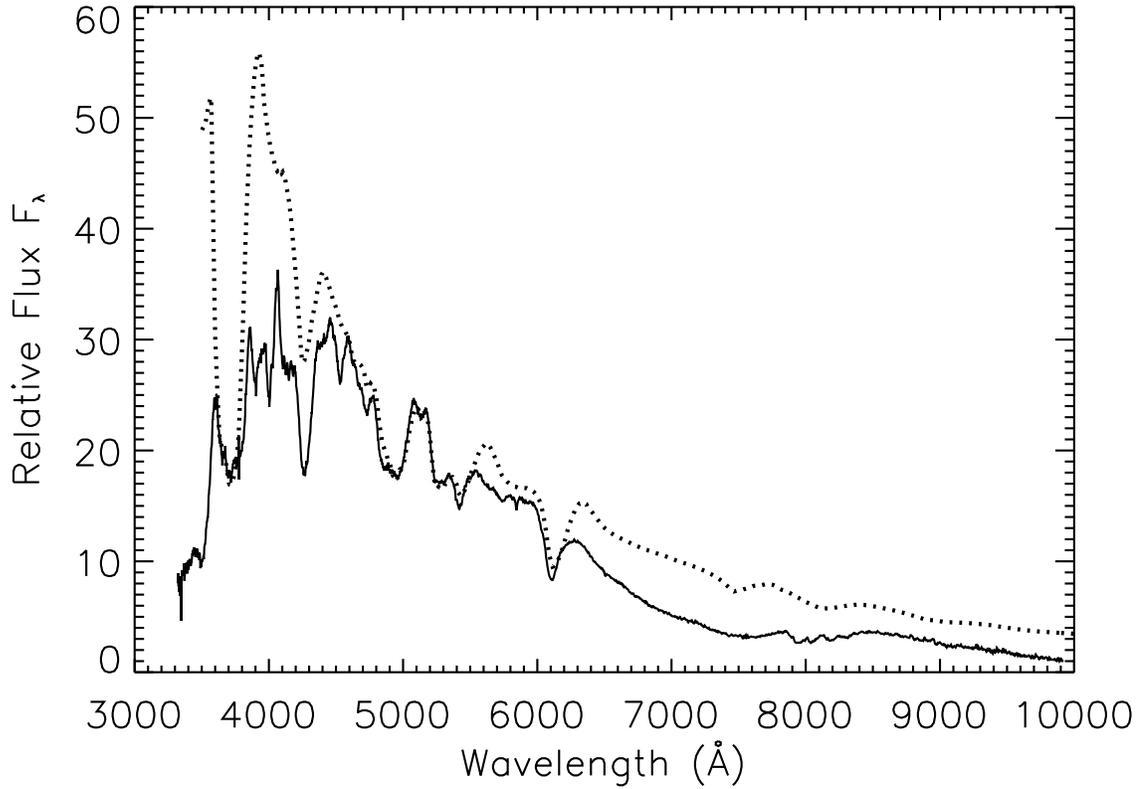}
\caption
{
 \Synow\ fit 1D1 (dotted line) of SN 2000cx (solid line) two days
 after maximum light with one \caii\ component.  Optical depths and
 excitation temperatures used for the other ions (\ion{Fe}{3},
 \ion{Si}{2}, \ion{S}{2}, and \ion{O}{1}) are as listed in
 Table \ref{tab:noCa}.
 \label{fig:1d1big}
}
\end{figure}

Here we present three illustrative fits of the \caii\ features using
only one spherically symmetric velocity component in each.  Figure
\ref{fig:1d1big} shows an example of a fit to the entire spectrum
range.  The observed features between 4500 \AA\ and 6000 \AA\ are fit
rather well assuming the parameters for \ion{Fe}{3}, \ion{Si}{2}, and
\ion{S}{2} given in Table \ref{tab:noCa}.

\begin{figure}
\figurenum{4}
\epsscale{0.74}
\plotone{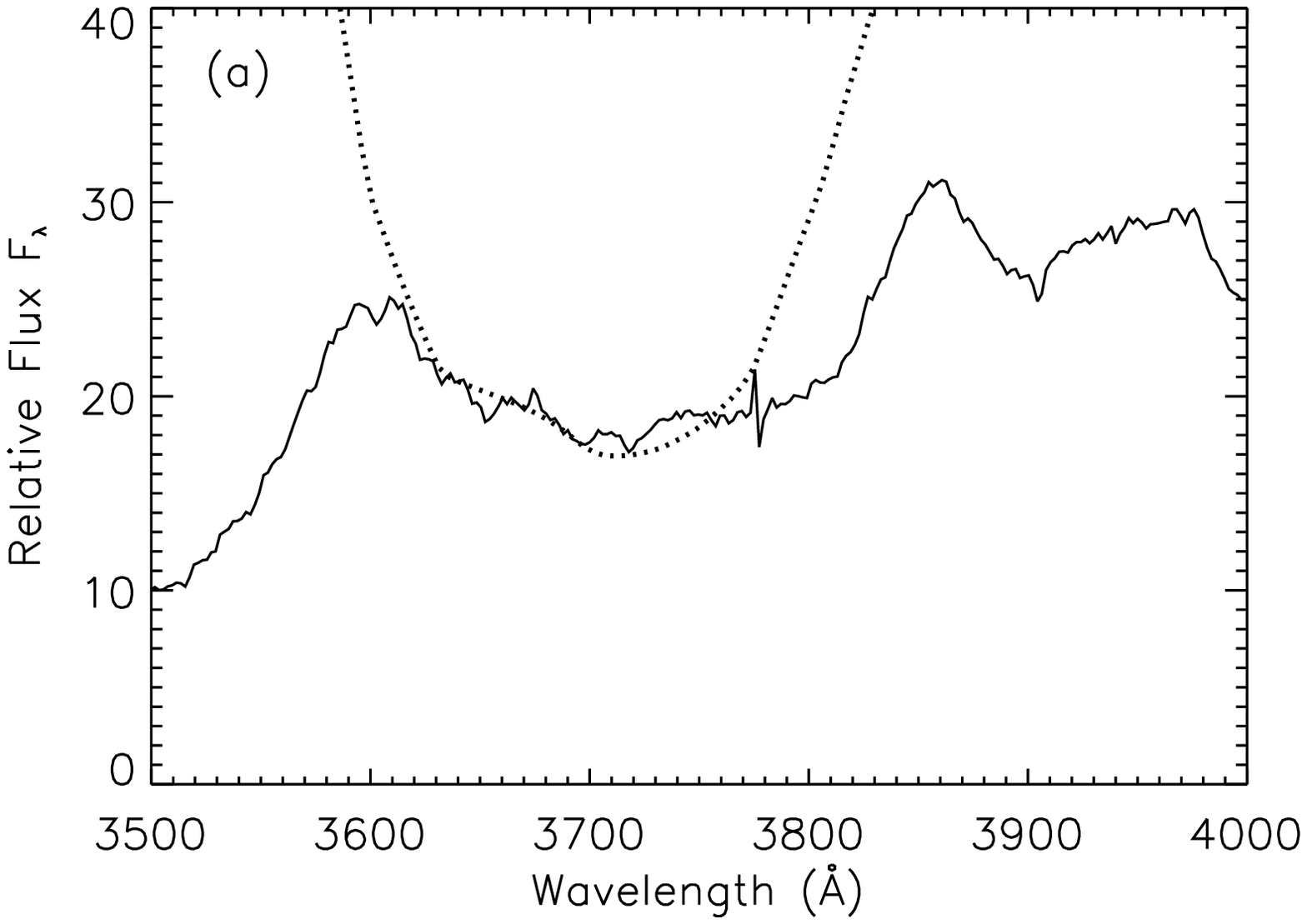}
\plotone{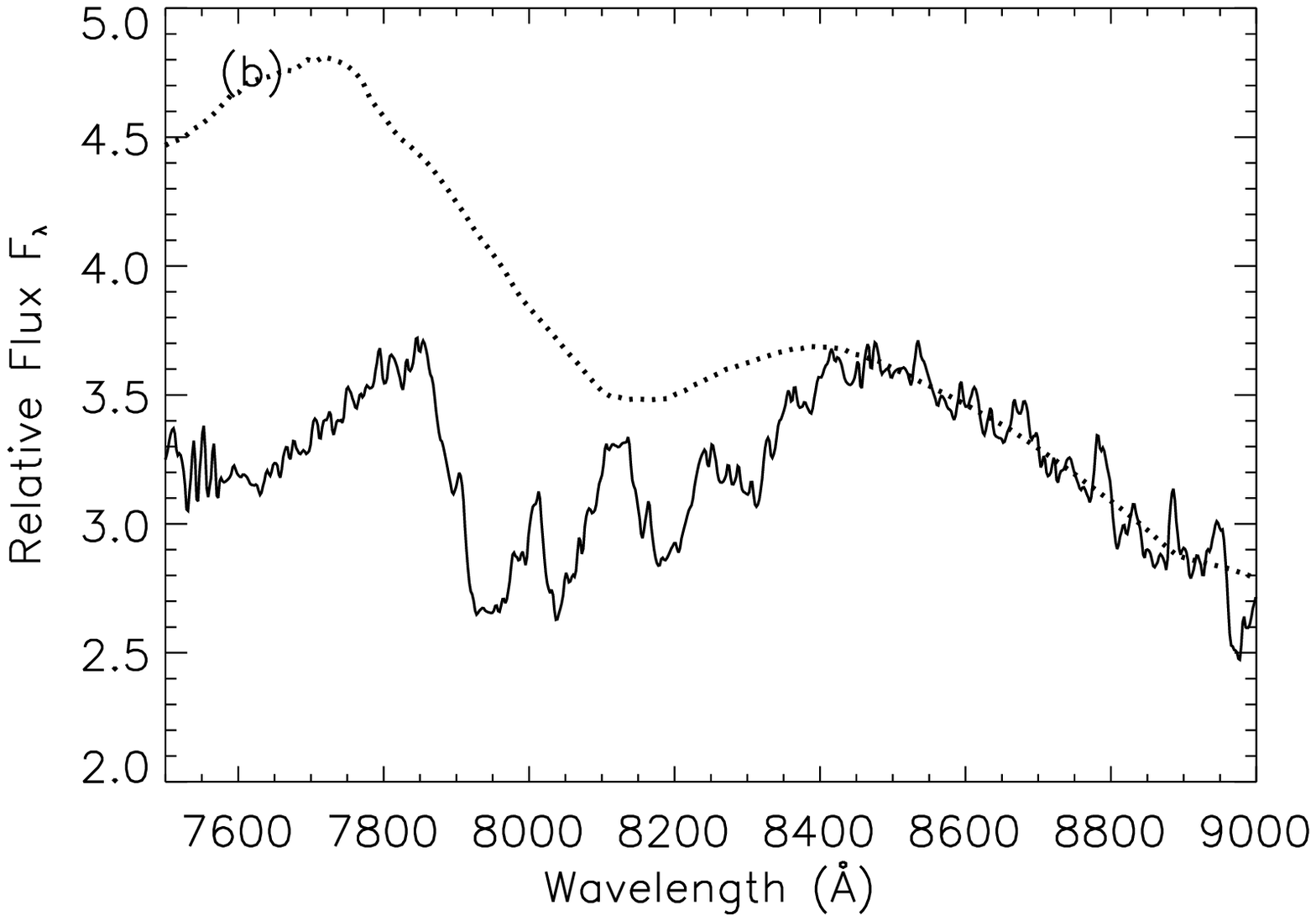}
\caption
{
 \Synow\ fit 1D1 (dotted line) of \caii\ features in SN 2000cx (solid 
 line) two days after maximum light.  A single, nearly constant optical
 depth shell extending from the photosphere to 30,000 \kmps\ is used.
 The velocity extent of \caii\ \handk\ is approximately reproduced, but the
 synthetic IR triplet lacks any of the structure seen in the
 observed spectrum.
 \label{fig:1d1}
}
\end{figure}

In Figure \ref{fig:1d1} are close-ups of the \handk\ feature and IR
triplet fits using model 1D1 from Table \ref{tab:synow}.  This single-shell
model with nearly constant optical depth as a function of radius
can reproduce the broad velocity extent of the \handk\ feature.  Yet
it cannot reproduce any of the structure present in the IR triplet, so
next we consider independent fits to the PV and HV features to investigate
this structure.

The appearance of the two strongest IR triplet lines (\caii\ \ww 8542,
8662) in the PV feature as two distinct notches presents an
interesting problem under spherical symmetry.  In general, without
assuming $v_{min} > v_{ph}$ or imposing a finite $v_{max}$,
generating such a feature in spherical symmetry is only possible if
$v_{ph} \lesssim v_{sep}$, where $v_{sep}$ is the velocity separation
of the two lines.  To fit the \ion{Si}{2} and \ion{Fe}{3} features,
$v_{ph} = 12,500$ \kmps\ is used, but $v_{sep}$ for the two strongest
IR triplet lines is only about 4000 \kmps\ and both notches have
minimum wavelengths consistent with a 12,500 \kmps\ blueshift.

\begin{figure}
\figurenum{5}
\epsscale{0.74}
\plotone{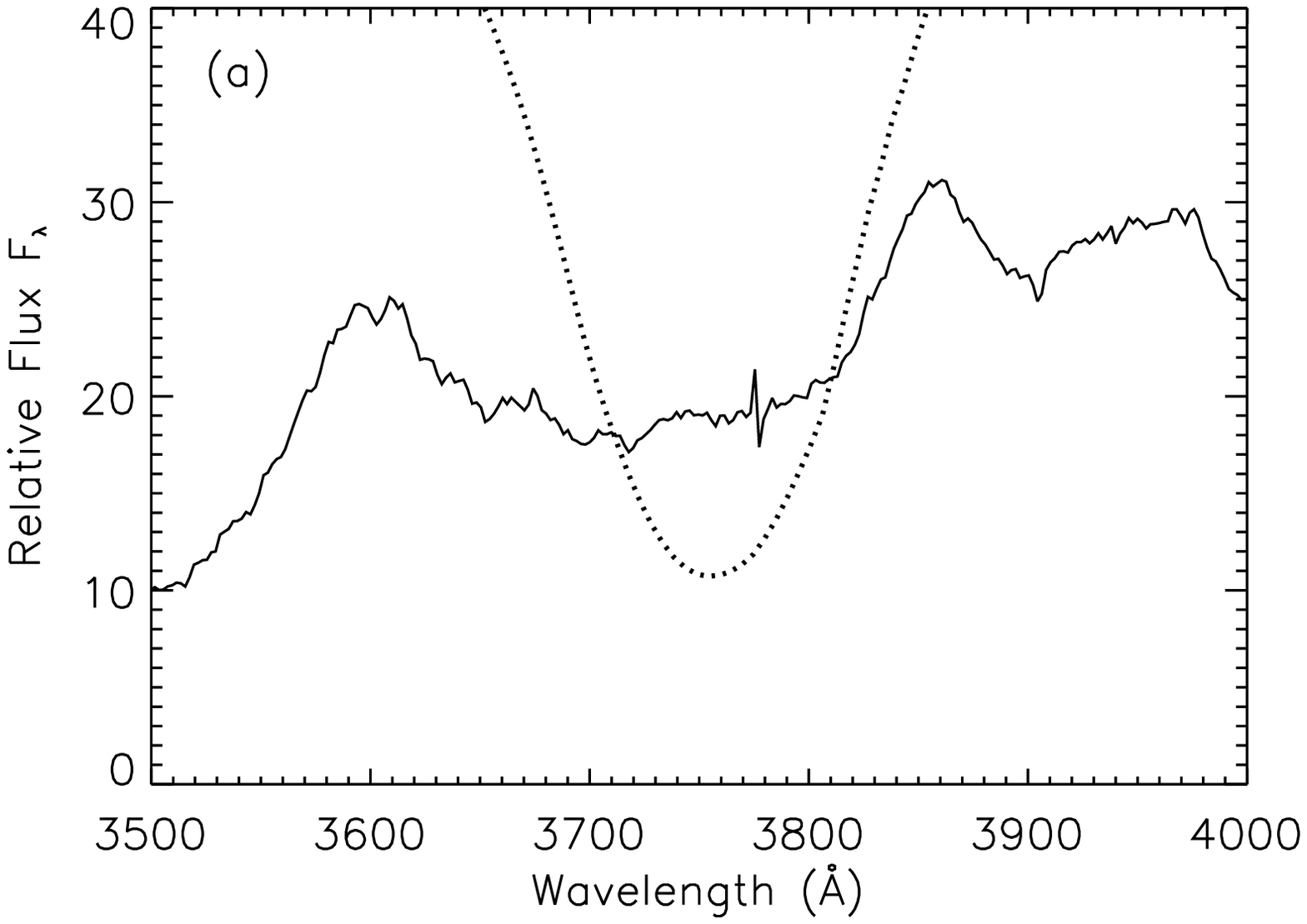}
\plotone{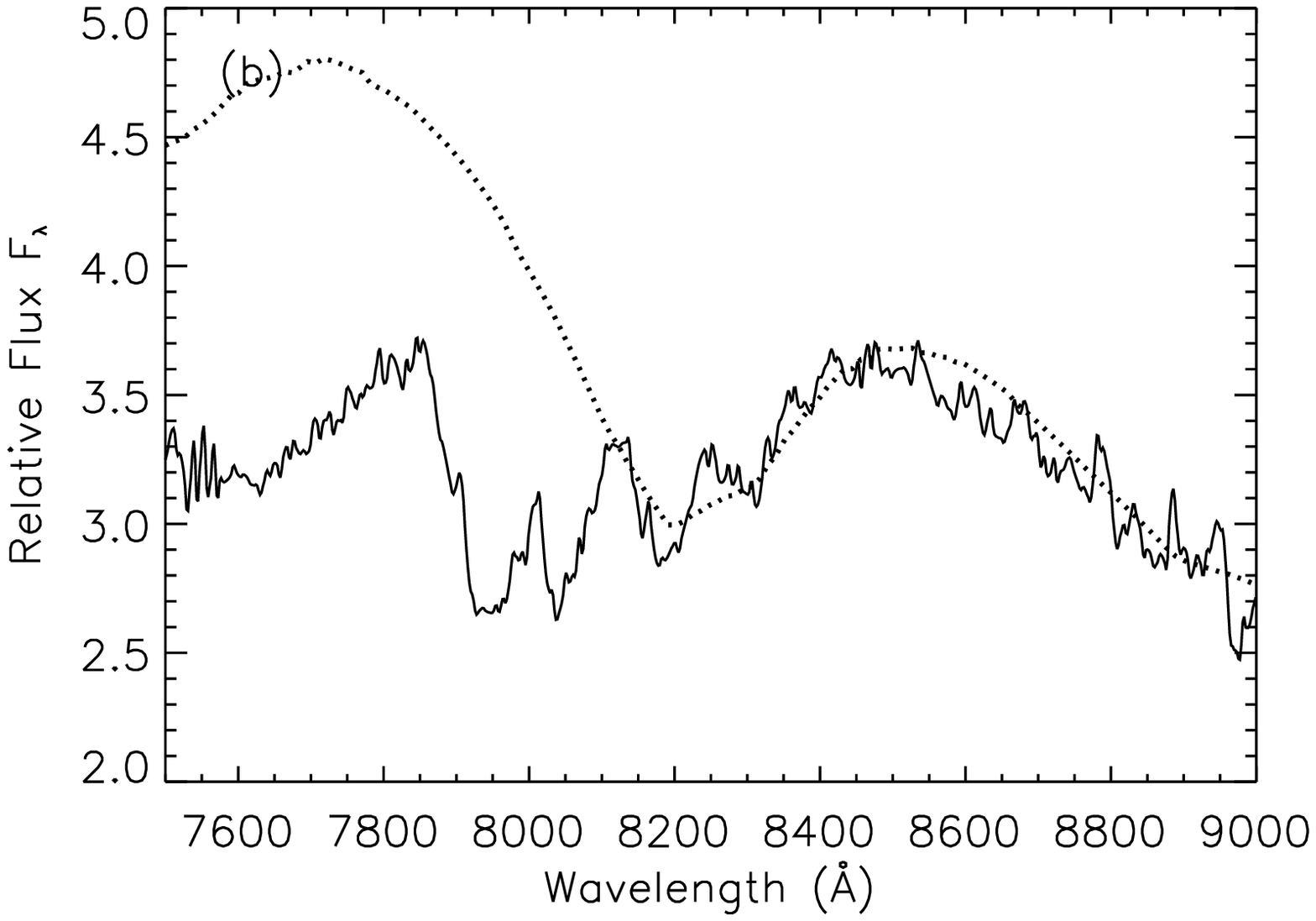}
\caption
{
 \Synow\ fit 1D1PV (dotted line) of \caii\ features in SN 2000cx (solid 
 line) two days after maximum light.  A single, exponentially decreasing
 optical depth component just above the photosphere is used.
 \label{fig:1d1pv}
}
\end{figure}

An alternative is to impose a finite $v_{max}$ or small $v_e$ that
prevents blending of the two absorptions into one.  But this has the
undesirable effect of making absorptions with flat bottoms which (when
combined together) generate a \caii\ \wl 8542 feature {\it shallower}
than the redder line, even though the former has a higher oscillator
strength.  For the moment, we allow the features to blend into one
absorption, using the parameters listed for fit 1D1PV in Table
\ref{tab:synow} which provide a satisfactory fit to the PV features
in Figure \ref{fig:1d1pv}.

\begin{figure}
\figurenum{6}
\epsscale{0.74}
\plotone{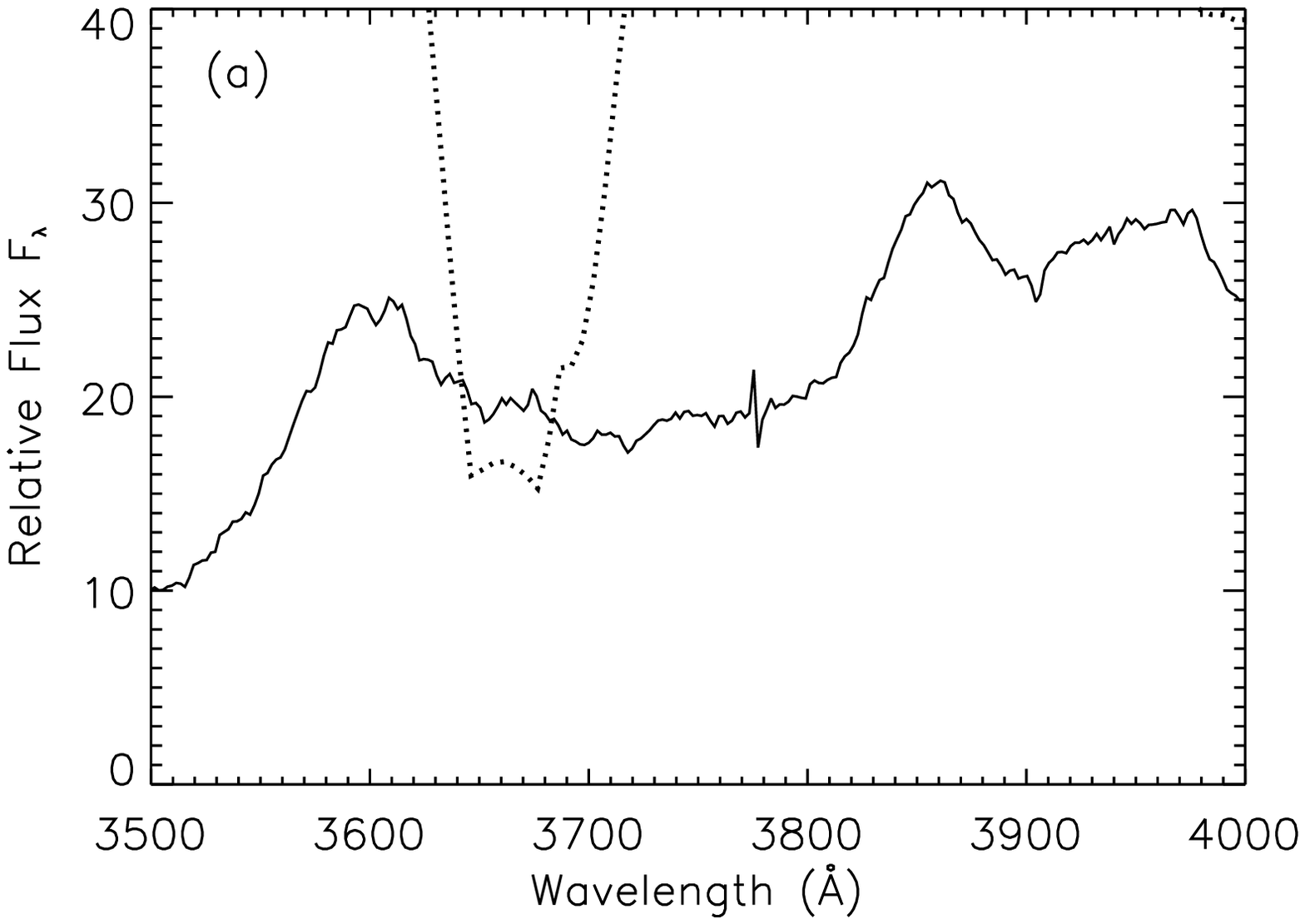}
\plotone{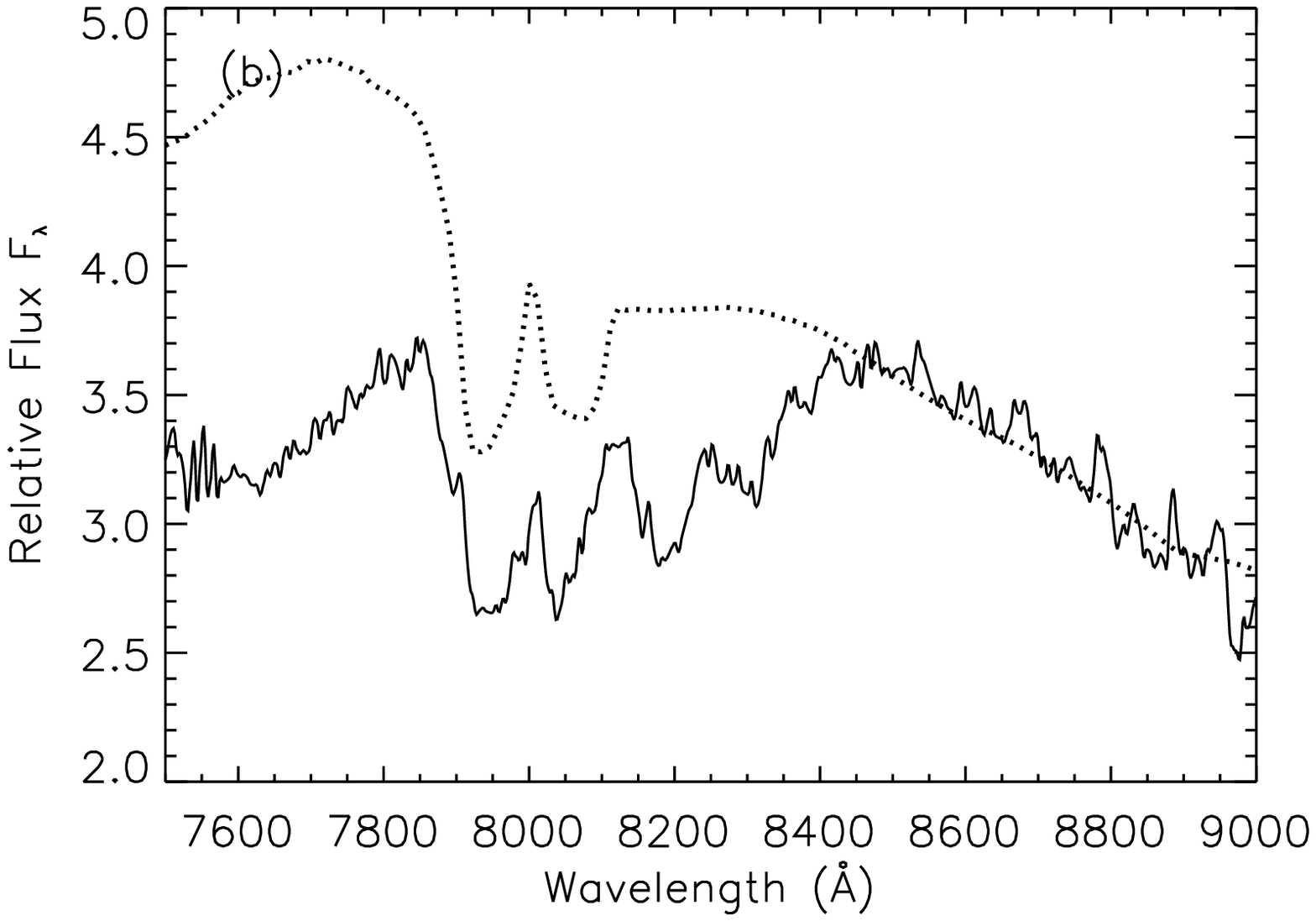}
\caption
{
 \Synow\ fit 1D1HV (dotted line) of \caii\ features in SN 2000cx (solid
 line) two days after maximum light.  A single, exponentially decreasing
 optical depth component at 20,000 \kmps\ is used.  An extremely
 small $e$-folding velocity of 500 \kmps\ is needed to prevent
 blending.
 \label{fig:1d1hv}
}
\end{figure}

Fitting the HV notches with a shell of \caii\ optical depth (fit 1D1HV
in Figure \ref{fig:1d1hv}) is considerably less problematic than in
the case of the PV feature.  Using a small $v_e$ or imposing a finite
$v_{max}$ prevents blending that would otherwise unite the two bluer
notches of the IR triplet feature.

\subsubsection{Two-Component Fit}

\begin{figure}
\figurenum{7}
\epsscale{0.74}
\plotone{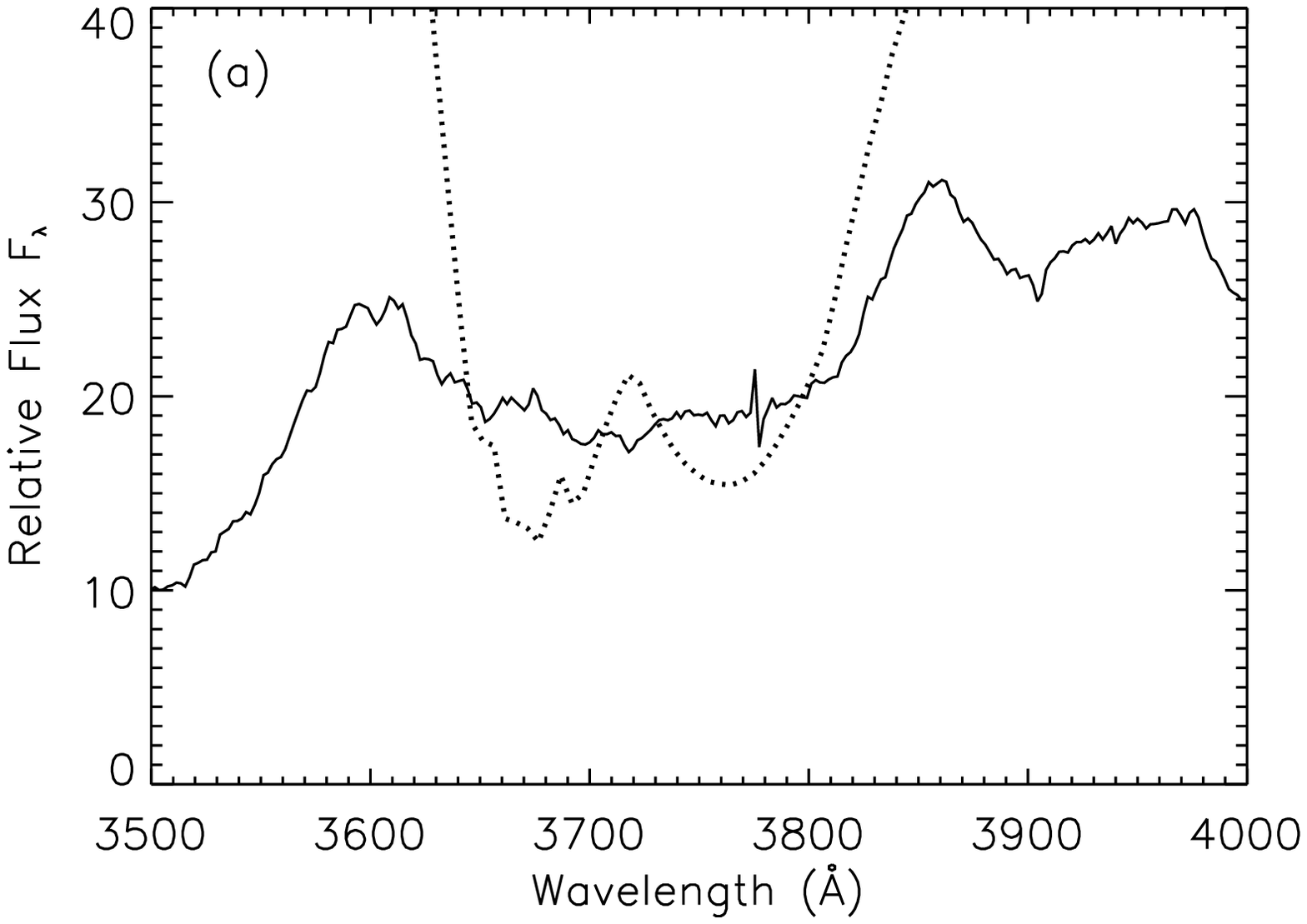}
\plotone{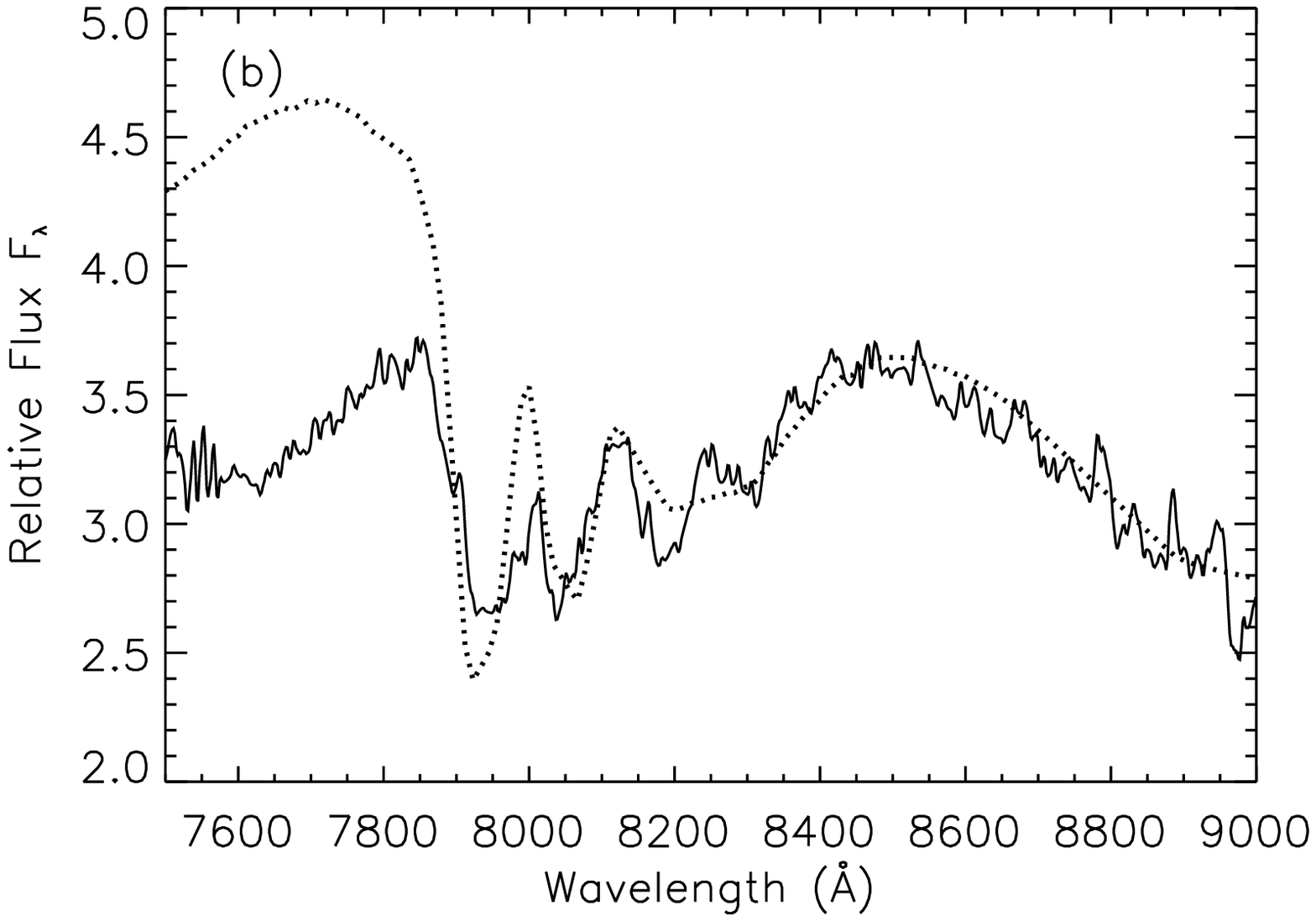}
\caption
{
 \Synow\ fit 1D2 (dotted line) of \caii\ features in SN 2000cx (solid 
 line) two days after maximum light.  Two exponentially decreasing shells of 
 \caii\ optical depth are used.  One extends from the photosphere to
 24,000 \kmps\ and the other from 24,000 \kmps\ to 25,000 \kmps.
 \label{fig:1d2}
}
\end{figure}

Combining the one-component PV and HV fits described above yields fit
1D2 (Figure \ref{fig:1d2}).  This fit is satisfactory for the IR
triplet, but its major deficiency is that the peak between the
two blue notches is higher than observed.  Adjusting $v_{max}$
from 25,000 \kmps\ to higher velocity permits the redder feature to
weaken this peak,
but extends the bluest synthetic notch further to the blue than is desired.
On the other hand, the synthetic \handk\ feature appears to need some
higher-velocity material to extend its blue edge.

\subsubsection{Three-Component Fit}

\begin{figure}
\figurenum{8}
\epsscale{0.74}
\plotone{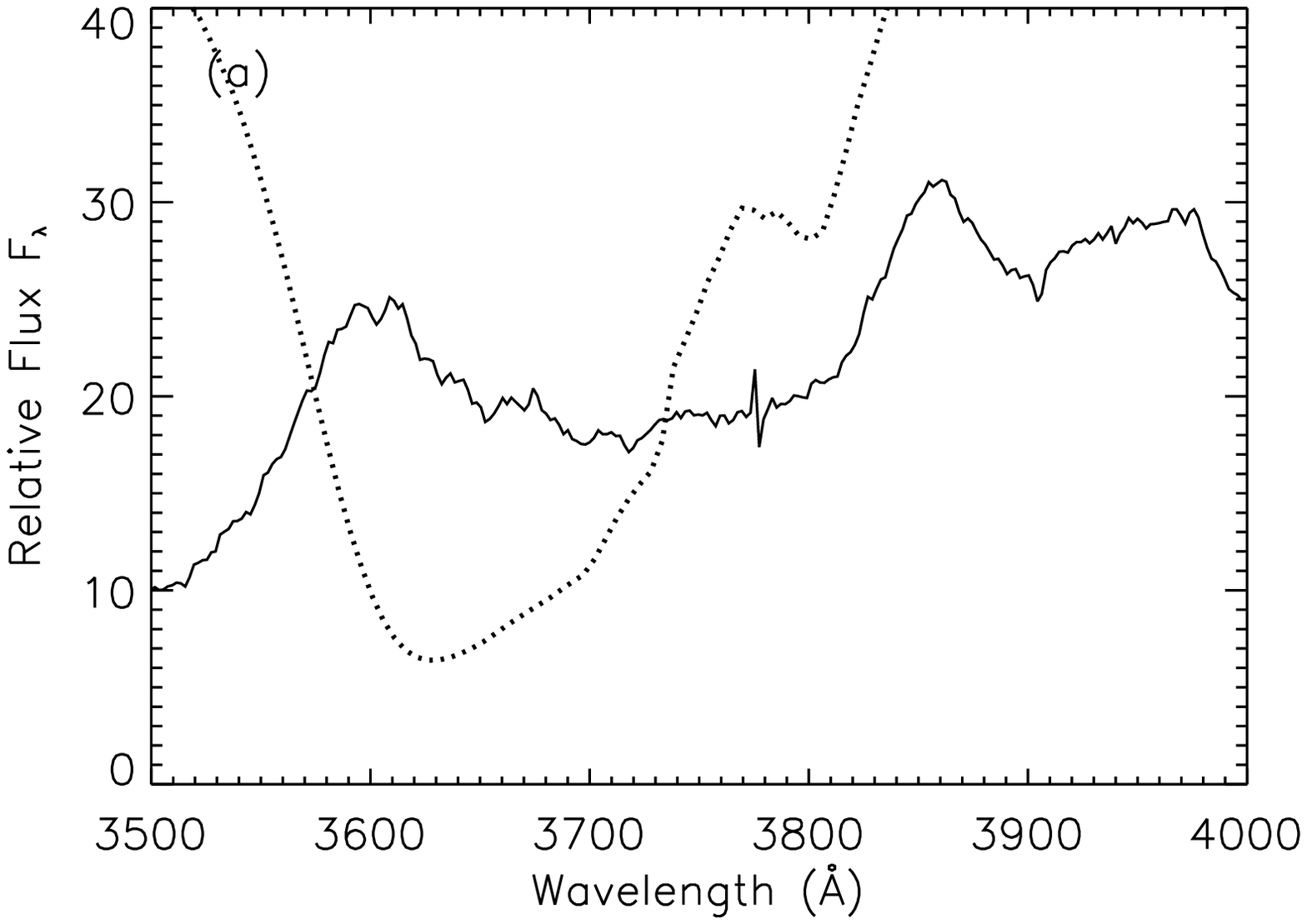}
\plotone{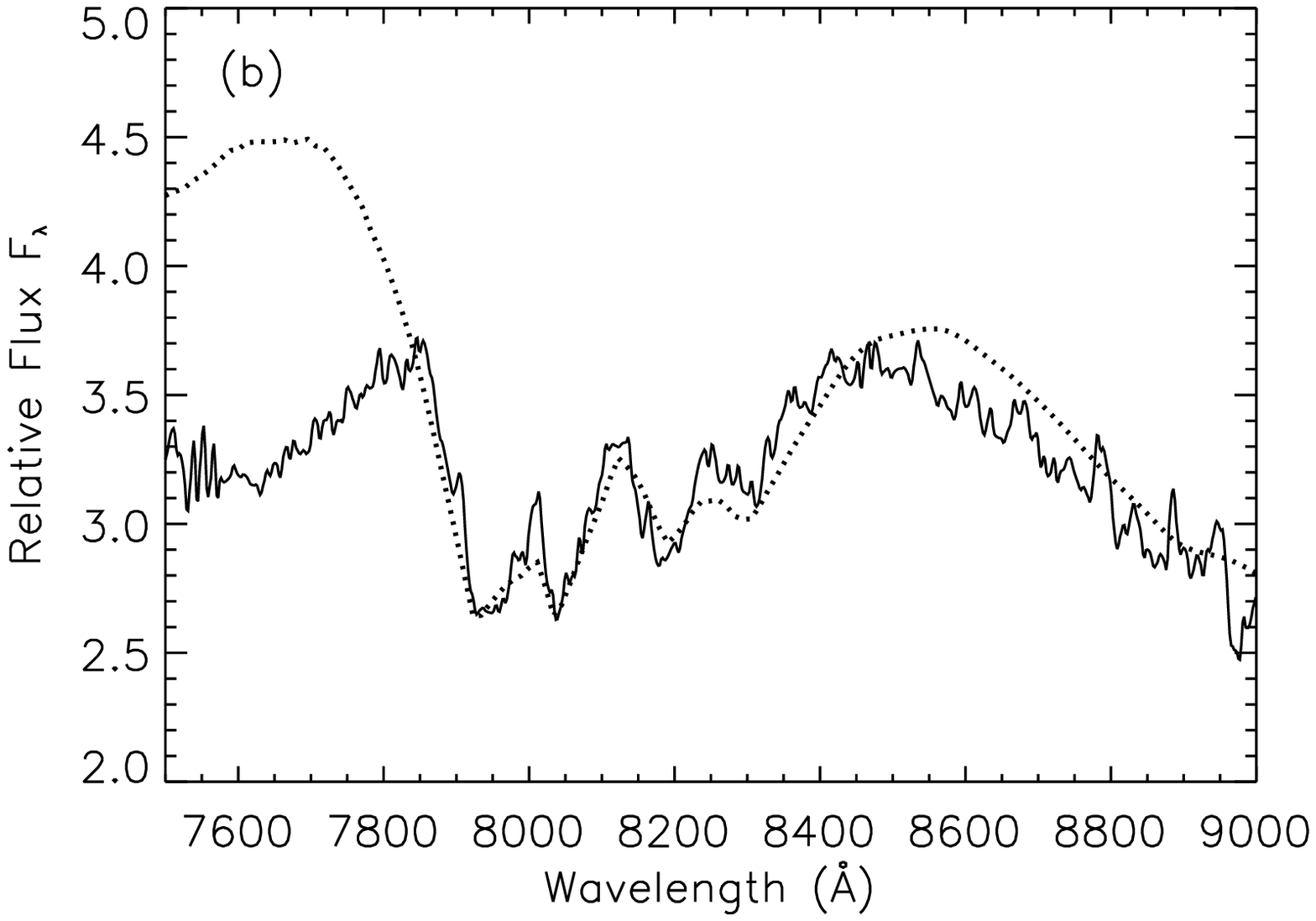}
\caption
{
 \Synow\ fit 1D3 (dotted line) of \caii\ features in SN 2000cx (solid 
 line) two days after maximum light.  Three exponentially decreasing shells of 
 \caii\ optical depth are used to generate the peak between the two red
 notches of the triplet.  The synthetic \handk\ feature is too strong 
 compared to the observed one.
 \label{fig:1d3}
}
\end{figure}

Using three velocity components of \caii\ optical depth allows us to
fit every notch in the IR triplet absorption by simply overlapping
pairs of notches formed by each component.  The result of this model
(1D3) is displayed in Figure \ref{fig:1d3}.  For the IR triplet, this appears
to be the ``best fit'' among the 1D models, but the \handk\ feature is
too strong in its bluest part.  Imposing a finite $v_{max}$ to
counteract this HV tail only deepens the synthetic \handk\ absorption 
since it removes material that scatters light from the emission lobes
of the envelope.

\subsection{\Brute\ Fit}

The assumptions used in \Brute\ are roughly the same as those in
\Synow\, except for the freedom to choose spatial optical depth
distributions without spherical symmetry.  Again the same blackbody
emitting, sharp photosphere is assumed, resulting in the same continuum-level
problem described before, and we use the same remedy here.
The technique used for radiative transfer in \Brute\ is that of Monte Carlo, a
difference from \Synow, but both codes take into account multiple
scattering in the calculation of the source functions of the lines.

The spatial parameterization used for the \caii\ optical depth consists
of two components.  One is spherical (PV material) and the other is
not (HV material).  The idea is to engineer a simple 3D distribution
of optical depth that yields a synthetic spectrum consistent with
observation, and fits at least as well as the 1D model.

\begin{figure}
\figurenum{9}
\epsscale{1.00}
\plotone{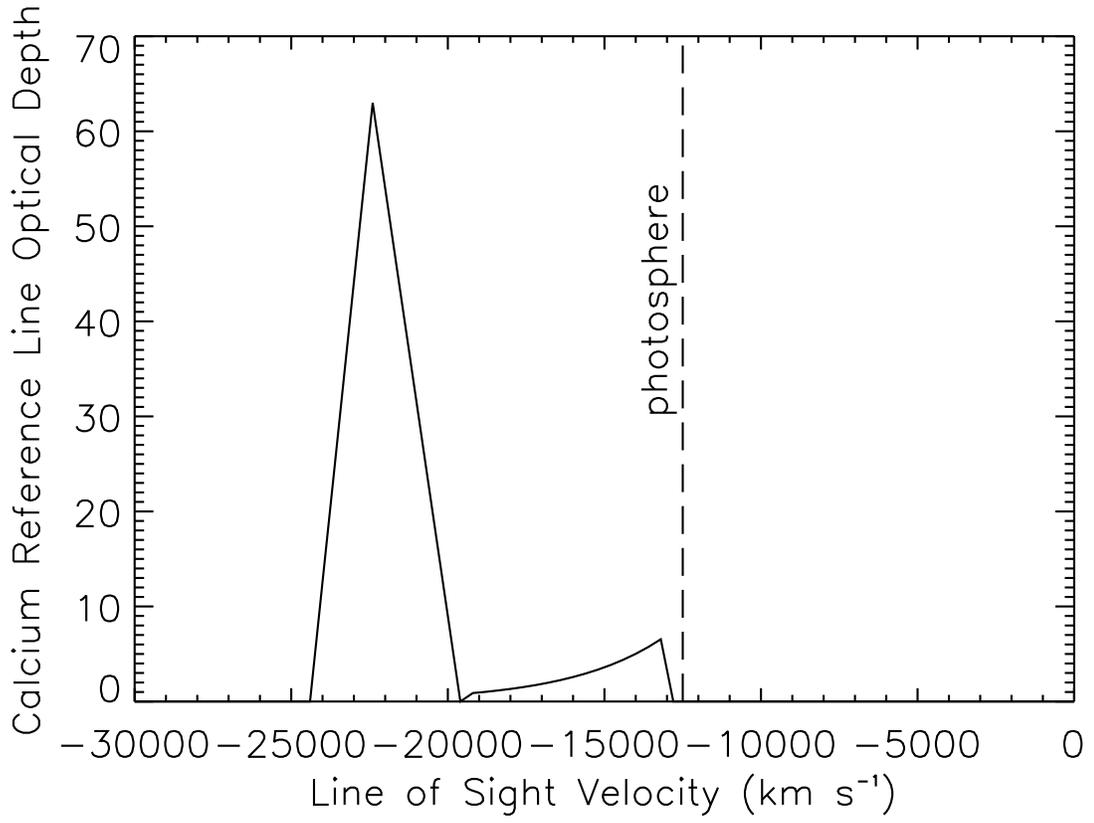}
\caption
{
 \caii\ reference line optical depth along the $-z$ axis in the 3D
 model.
 \label{fig:tau}
}
\end{figure}

For the HV material in the 3D fit, we adopt a simple geometry
consisting of a circular cylinder of radius 8000 \kmps\ coaxial with
the line of sight to the center of the photosphere.  The reference
optical depth for \caii\ inside the cylinder is assigned according 
to the rule
\begin{equation}
 \tau_{ref}(v_z) =
 \left\{ 
 \begin{array}{r@{\quad:\quad}ll}
  0.0225 v_z - 441.0 &
  19,600 < v_z < 22,400 & \textrm{\kmps} \\
  -0.0315 v_z + 768.6 & 
  22,400 < v_z < 24,400 & \textrm{\kmps}.
 \end{array} \right.
 \label{eq:cylRule}
\end{equation}
According to this rule, the optical depth in the cylinder reaches a
maximum value of $\tau_{ref} = 63$ at the plane $v_z = 22,400$ \kmps.
Outside the cylinder, the optical depth for the \caii\ reference line is
prescribed by eq. (\ref{eq:sphRule}) with $\tau_{ref}(v_{min}) =
7$, $v_e = 3000$ \kmps, and $v_{min} = 13,000$ \kmps.  A plot of
reference optical depth along the line of sight is presented in Figure
\ref{fig:tau}.  The reference optical depths of the other ions are
parameterized exactly as in the \Synow\ fits and produce the same
features.

\begin{figure}
\figurenum{10}
\epsscale{0.74}
\plotone{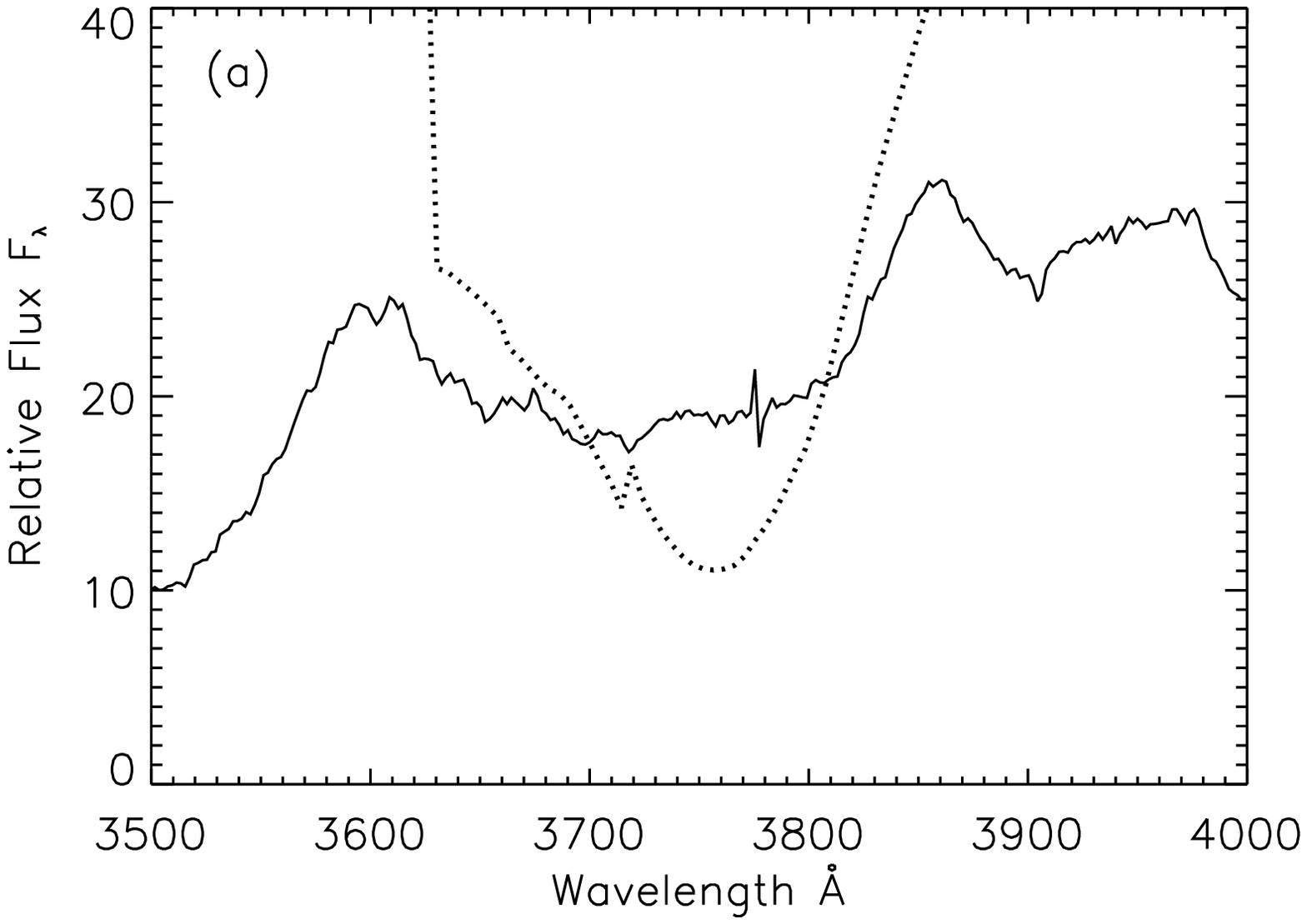}
\plotone{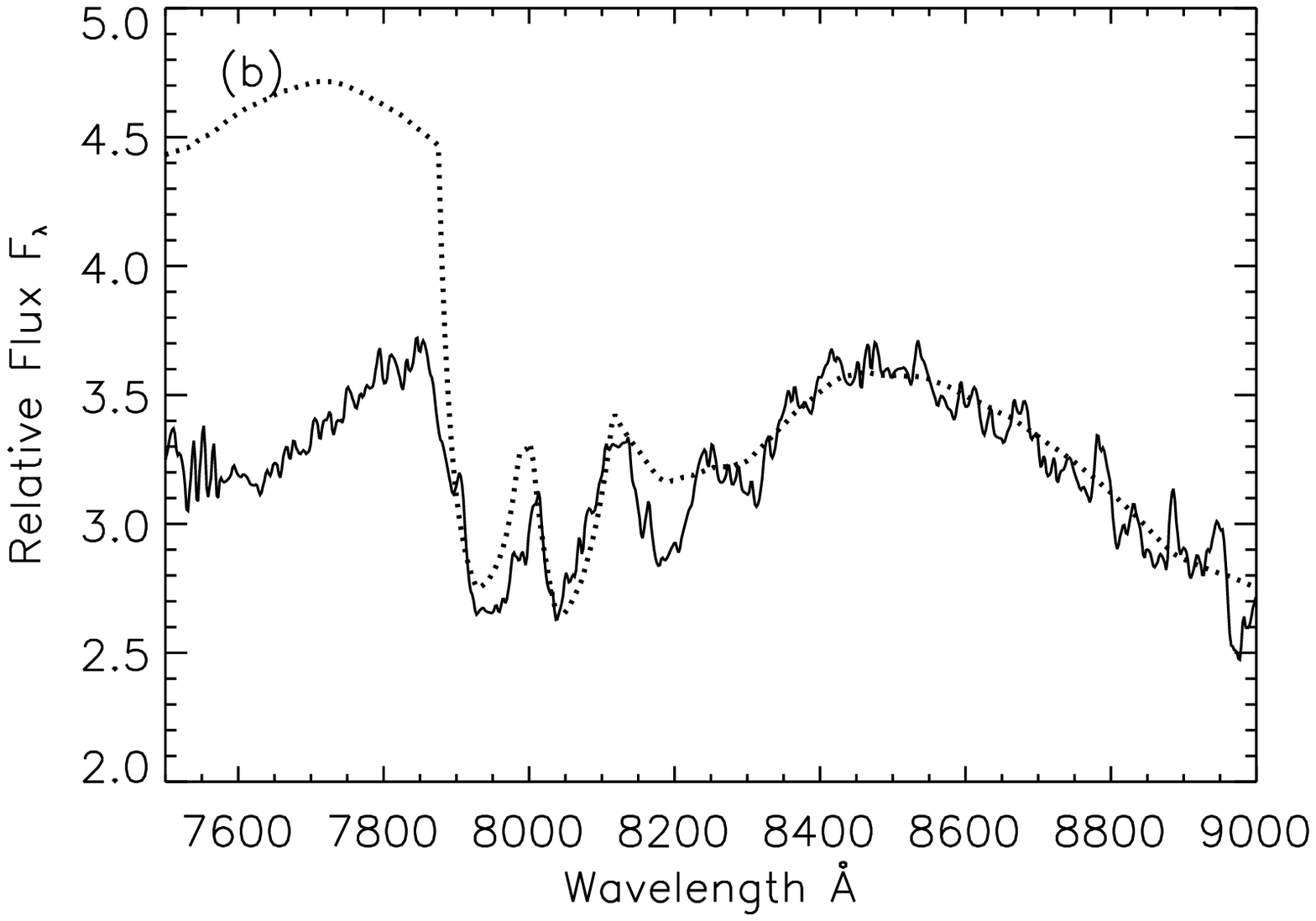}
\caption
{
 \Brute\ fit (dotted line) of \caii\ features in SN 2000cx (solid 
 line) two days after maximum light. The IR triplet is fit, and the blue
 edge of the synthetic \handk\ feature is weakened with respect to fit 1D3.
 \label{fig:3d}
}
\end{figure}

The \caii\ features resulting from this 3D parameterization are shown
in Figure \ref{fig:3d}.  The fit to the HV IR triplet is rather good.
The synthetic HV \handk\ is more consistent with the observation than
in the 1D2 or 1D3 parameterizations.  Since the PV \caii\ optical
depth is distributed much the same as in the 1D2 case, the weak observed
peak between the redder notches cannot be reproduced.  In the 1D case,
this could only be accomplished by adding a third shell of optical
depth and tuning its position.  We note that in the 3D case, if the PV
material is deployed in a nonspherical manner (such that parts of the
photosphere shine through), then the peak between the two PV notches
can be reproduced quite easily.


\section{Discussion}

\subsection{HV Ejecta Geometry from the Fits}

Reconstructing SN envelope structure from spectra is an ill-posed
inverse problem.  The problem becomes even more difficult when the
assumption of spherical symmetry is relaxed; the presented 3D
solution represents but one of many possible solutions.

In fact, to first order, any HV distribution of \caii\ optical 
depth which yields the
same photospheric covering factor as a function of $v_z$ will produce
the same absorption features.  The most promising avenue for breaking
this degeneracy is through synthetic spectropolarimetry of high-quality
data.  The Stokes polarization parameters Q and U for
wavelengths covering the \caii\ features are required.  Such Q-U plots
are the means by which the HV ejecta geometry of SN 2001el is
constrained \citep{Kasen2003}.

Both that study and the present one focus on the absorption
features of \caii, and ignore the emission features.  Generally,
emission phenomena are of limited utility in assessing HV ejecta
geometry.  In spherical symmetry, a HV shell gives rise to a weak,
flat-topped emission feature barely detectable in flux spectra.  On
the other hand, if the HV ejecta are organized into isolated regions
of high optical depth, those not covering the photosphere along the
line of sight will only contribute individual emission bumps to the
spectrum.  These weak features are even harder to detect if the PV
ejecta are spherically symmetric and contribute significantly to the
emission feature.

On the practical side, the peculiarities of the spectrum of SN 2000cx
make it impossible to use the \caii\ emission features for geometrical
analysis.  The \handk\ emission feature is disrupted by a sequence of
narrow lines, while the IR triplet emission is plagued by fringing 
in the CCD.

These factors (the inherent degeneracies of the problem, the lack of
corroborating polarization data, the limited utility of emission
features) prevent us from making all but the most conservative
statements about the HV ejecta geometry of SN 2000cx.
The similarity of the HV IR triplet of SN 2000cx to that of SN 2001el
is suggestive, and its geometrical interpretation in the latter event
motivates our consideration of a 3D model for SN 2000cx.

From a purely empirical standpoint, none of the presented fits stands
out as the ``best'' overall.  Any cosmetic details missed by the
synthetic spectra could be tuned away by making minor adjustments to
the models.  However, the main goal is to simultaneously fit the HV
\handk\ and IR triplet, and the 3D fit seems to do a better job.  

Regardless of the details of the true geometry of the HV \caii\ in SN
2000cx, the presence of this feature in only some SNe Ia evokes a
series of questions to address in the future as high-quality data
become available and 3D spectrum analysis techniques improve.

(1) How frequent is HV ejecta clumping in SNe Ia?  All SNe Ia might
possess HV deviation from spherical symmetry which could escape
detection if optically thick parts did not obscure the photosphere.
SNe 2001el and 2000cx might be examples where fortuitous orientation of
HV ejecta permitted detection.  This explanation is bolstered by the
otherwise normal appearance of the SN 2001el spectrum.  However, the
peculiarities of SN 2000cx \citep[the \ion{Fe}{3} lines in its
spectrum, its unusual light curve and color evolution;][]{Li2001}
make this hypothesis somewhat problematic.

(2) What implication does the presence of clumpy HV ejecta have for 
explosion models?  If some SNe Ia have nonspherical HV ejecta while
others do not, it might imply that more than one progenitor model
is needed to explain the SN Ia phenomenon.  More conservatively, the
clumpiness of the HV ejecta might be a consequence of a so-called
second parameter \citep{Branch2001}.

(3) Does the existence of nonspherical SNe Ia influence the prospects
for precision cosmology?  The most obvious spectroscopic effects would
be due to the application of template-derived K-corrections
\citep{Nugent2002} from spherical SNe to nonspherical ones.  The
effect would be minor when the spectroscopic outcome of nonsphericity
is isolated to a few lines (e.g. \ion{Si}{2}).  On the other hand,
line-blanketing ions (e.g. \ion{Fe}{2}) will greatly influence the
photometry along some lines of sight more than others.

\caii\ lines are notoriously excitable over a large range of
temperature and density.  This makes it possible to place limits on
the HV ejecta mass, and in turn to constrain future white dwarf explosion
models.

\subsection{HV Ejecta Mass Estimates}

Here we make some simple estimates of the HV ejecta mass as
constrained by the \caii\ features visible in the SN 2000cx spectrum
at two days after maximum light.  We make two pairs of estimates,
each pair consisting of a 1D and 3D measurement.  The first pair of
estimates is made for the extreme case of a purely calcium HV ejecta 
composition.  The second pair is based on a C/O-rich composition,
similar to the outer, unburned layers of a white dwarf.  The details
of this composition are as listed in the SN ion signatures atlas of
\citet{Hatano1999b}.

For each pair of estimates, we adopt the 1D2 and 3D parameterization
domains for the volume used in calculating the HV ejecta mass:
\begin{equation}
 \begin{array}{ccll}
  V_{HV}^{1D2} & = & 3.9 \times 10^{46} & \textrm{\cc}, \\
  V_{HV}^{3D}  & = & 5.0 \times 10^{45} & \textrm{\cc}.
 \end{array}
\end{equation}
These presented mass estimates are not intended to be exact; in all 
cases the assumption of thermal equilibrium (TE) is 
employed.  More rigorous constraints
on the HV ejecta mass require detailed modelling of the physical 
conditions in the HV material, and these will be the focus of much future work.
The estimates here and in \citet{Kasen2003} are a starting point
for that effort.

\subsubsection{Pure Calcium Composition}

In the Sobolev approximation, line optical depth for a transition $l
\rightarrow u$ in a SN envelope is given by \citep{Jeffery1990}
\begin{equation}
 \tau_{lu} = \frac{\pi e^2}{m_e c}
             f_{lu} \lambda_{lu} n_l t \biggl( 1 - \frac{g_l n_u}{g_u
             n_l} \biggr),
 \label{eq:sobolev1}
\end{equation}
where $f_{lu}$ is the oscillator strength, $\lambda_{lu}$ is the
transition wavelength, $n_l$ and $n_u$ are the lower and upper level
occupation number densities, respectively, and $t$ is the time 
since explosion.  The
other symbols have their usual meanings.  For convenience, we rewrite
this as
\begin{equation}
 \tau_{lu} = 2.292 \times 10^{-5} (gf)_{lu} \lamang n t_d
             \frac{\exp( - E_l / kT )}{Q(T)},
 \label{eq:sobolev2}
\end{equation}
where we have neglected the correction factor for stimulated emission,
replaced the constants with their numerical values, and imposed TE.
The partition function $Q(T)$ for the ion in question is
included.

If the HV ejecta consist of singly ionized calcium in the ground
state, then the optical depth in the \caii\ \wl 3934 transition (the
reference line) becomes roughly proportional to the mass density of
the HV ejecta.  In reality, the gas consists of other ions of calcium
and other species, so deriving a mass density assuming a pure \caii\
composition provides an extreme lower limit to the HV mass.  
Substituting $\lamang = 3934$, $(gf)_{3934} = 1.3614$, $E_l = 0$ eV,
and $t_d = 20$ into eq. (\ref{eq:sobolev2}) gives
\begin{equation}
 \tau_{3934} \simeq 2.46 n_{Ca II} / Q(T).
\end{equation}
If we assume that the temperature in the HV ejecta is less than 8000 K, we can
use the simple approximation $\tau_{3934} \simeq n_{Ca II}$ since 
$Q(T)$ varies slowly from 2 to 3 up to 8000 K.  

Now an average mass density for the HV ejecta as derived from the 
\caii\ optical depth can be written
\begin{equation}
 \begin{array}{cclcll}
  \langle \rho_{HV} \rangle 
  & \simeq 
  & (A_{Ca}/N_{Avo}) n_{Ca}
  & \simeq 
  & 6.66 \times 10^{-23} \langle \tau_{HV} \rangle
  & \textrm{\gcc.}
 \end{array}
 \label{eq:rho}
\end{equation}
If we average the reference line optical depths over the 1D2 and 3D
parameterization domains, then 
$\langle \tau_{HV}^{1D2} \rangle = 25.5$ and 
$\langle \tau_{HV}^{3D}  \rangle = 31.5$.  Substituting these values
into eq. (\ref{eq:rho}) and multiplying by the volume in each
case, 
\begin{equation}
 \begin{array}{ccccc}
    M_{HV}^{1D2} 
  & \gtrsim
  & 3.3 \times 10^{-8} M_\odot, \\
    M_{HV}^{3D} 
  & \gtrsim
  & 5.2 \times 10^{-9} M_\odot. \\
 \end{array}
\end{equation}
These mass estimates are the ``rock bottom'' numbers required for the 
observational signature of the \caii\ \wl 8542 line.  The deflagration model
W7 \citep{Nomoto1984, Branch1985} suggests that the amount of material
and the densities in this region should be much higher, on the order of
$6.0 \times 10^{-3} M_\odot$ in spherical symmetry above 20,000 \kmps.
Including other species into the mass calculations improves the lower
limit, but this requires a choice of a particular composition.

\subsubsection{C/O-Rich Composition}

A more realistic composition model than that of pure calcium can be
used to produce a more meaningful constraint on the HV ejecta mass.
Here we adopt the C/O-rich composition from the ion signature atlas of
\citet{Hatano1999b} as the candidate model.  This composition is
representative of the unburned outer layers of an exploded white
dwarf.

The procedure for constraining the HV mass density is quite simple.
For a given temperature $T$ and an atomic line, we iteratively solve the
equation of state for the mass density $\rho_{tot}$ where $\tau = 1$ 
in the line.
Repeating this procedure for a range of $T$ yields a curve in
the $\rho_{tot}-T$ plane.  Generally, the space above the curve
represents densities for which the line optical depth exceeds unity.
In principle, if a line has a spectroscopic signature, then $\tau > 1$
in the line and the corresponding $\rho_{tot}-T$ curve gives a lower 
limit to the mass density.  Of course, this procedure depends on the
assumption of thermal equilibrium, so for some ions the curves are not
representative.  

\begin{figure}
\figurenum{11}
\epsscale{1.000}
\plotone{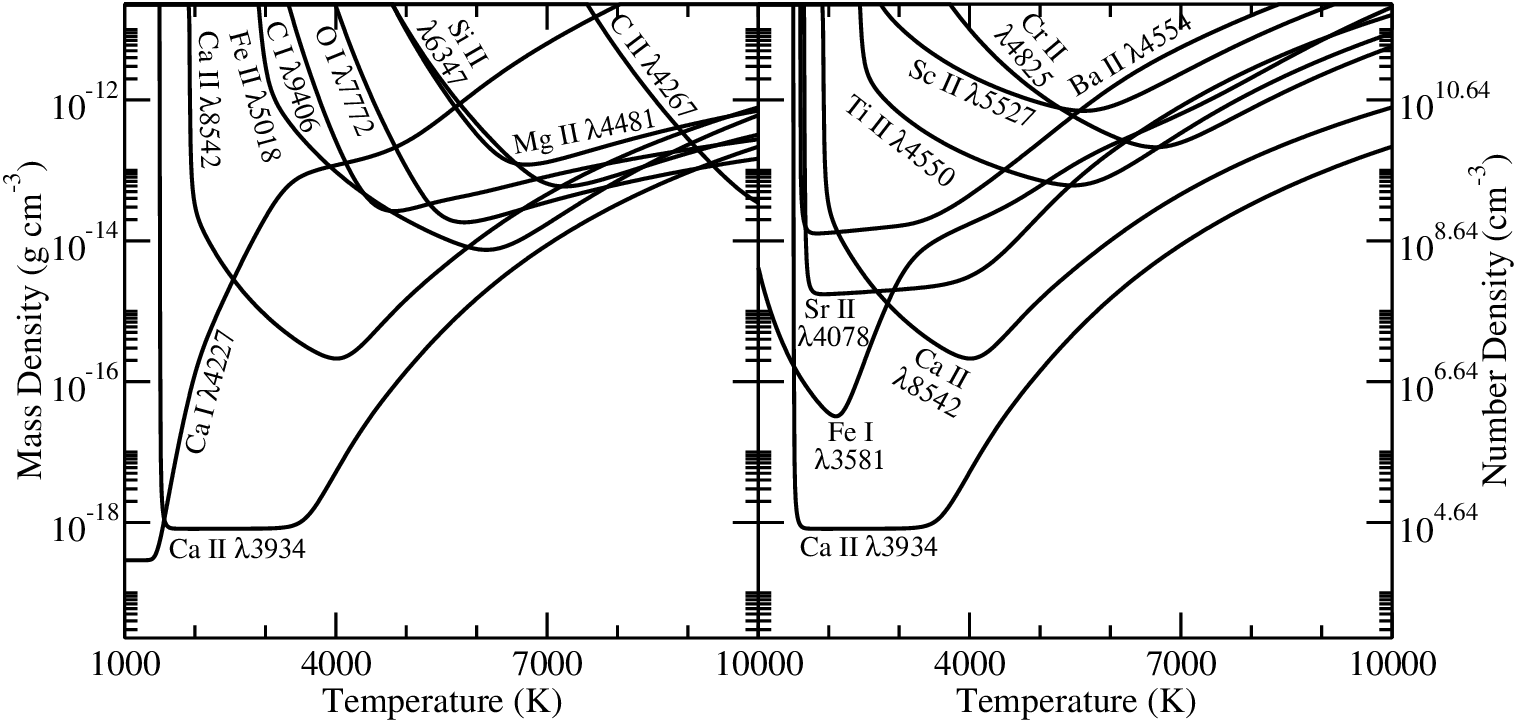}
\caption
{
 Contours of $\tau = 1$ for a variety of ions in the C/O-rich
 composition \citep{Hatano1999b}, assuming TE.  The \caii\ signature
 curves are repeated in both panels for comparison with curves of
 other ions.  
 \label{fig:refcont_co-rich}
}
\end{figure}

Contours of $\tau = 1$ for \caii\ \wl 3934 and \wl 8542 are shown
along with curves for reference lines of other ions in Figure
\ref{fig:refcont_co-rich}.  For the C/O-rich composition used, the
absolute minimum mass for which \caii\ \wl 8542 has a spectroscopic
signature is at $\rho_{min}^{C/O} = 2.2 \times 10^{-16}$ \gcc.

The value of $\rho_{min}^{C/O}$ depends on the choice of $T$, as does
the relative strength of \caii\ \wl 8542 to \wl 3934.  However, since
the optical depth of \caii\ \wl 3934 is always much greater than that
of the IR line, it is difficult to constrain the exact value of $T$ in
the HV ejecta using this technique.  As $\tau$ in a line approaches
$\infty$, the line profile saturates and the dependence of the absorption
feature depth on $\tau$ disappears.  Rather than resort to predictions
from detailed models for a temperature in the HV ejecta \citep[as was
done for SN 2001el by][]{Kasen2003}, we adopt the absolute minimum
value given above for a conservative estimate of the HV ejecta masses:
\begin{equation}
 \begin{array}{ccccc}
    M_{HV}^{1D2} 
  & \gtrsim
  & 4.3 \times 10^{-3} M_\odot, \\
    M_{HV}^{3D} 
  & \gtrsim
  & 5.5 \times 10^{-4} M_\odot. \\
 \end{array}
 \label{eq:co_est}
\end{equation}
These mass estimates and densities are roughly consistent with the
W7 estimate of $6.0 \times 10^{-3} M_\odot$ in spherical symmetry 
above 20,000 \kmps.  This might suggest that the \ion{Ca}{2} signature 
could arise from primordial material as was suggested for SN 1994D
\citep{Hatano1999a}, rather than from freshly synthesized material.

Other ion signature curves are plotted in Figure \ref{fig:refcont_co-rich}
along with the two \ion{Ca}{2} curves.  Portions of those curves
lying above the \ion{Ca}{2} \wl 8542 curve indicate candidate ions to
look for in the HV material.  Of particular interest is the
\ion{Fe}{2} curve: \citet{Hatano1999a} detected the presence of
\ion{Fe}{2} in SN 1994D at HV.  Another interesting ion is
\ion{Ti}{2}, usually detected as a trough of lines in SN 1991bg-like
events
\citep{Filippenko1992}.  
The notches just to the blue of the 4300 \AA\ \ion{Fe}{3}
feature could be attributed to HV \ion{Ti}{2}.  

One might consider using the non-detection of features of certain ions
to place an upper limit on the mass at HV.  This is difficult since
many of the lines are blended with other, stronger features.  For
example, HV \ion{Sr}{2} blends with the \caii\ \handk\ feature.  Lines
from \ion{Fe}{2}, especially if weak, blend with the \ion{Fe}{3}
features.  Since we cannot be certain these lines are present, though
weak and blended, we refrain from capping the HV mass estimate.

\subsubsection{Other Compositions}

\citet{Marietta2000} present interesting simulations of the effect
that a white dwarf in a binary has on its companion when it explodes.
Generally, they note that more evolved companions (with less tightly
bound envelopes) are more vulnerable to losing a substantial fraction
of their envelope during the explosion.  The resulting ejecta
distribution includes an evacuated cone in the ejecta behind the
companion and some hydrogen with characteristic velocity on the order
of 1000 \kmps.  However, a small amount of stripped hydrogen (about
$10^{-4} M_\odot$) could be carried up to velocities exceeding 15,000
\kmps.  Could the HV \caii\ lines be a signature of such material?

\begin{figure}
\figurenum{12}
\epsscale{1.000}
\plotone{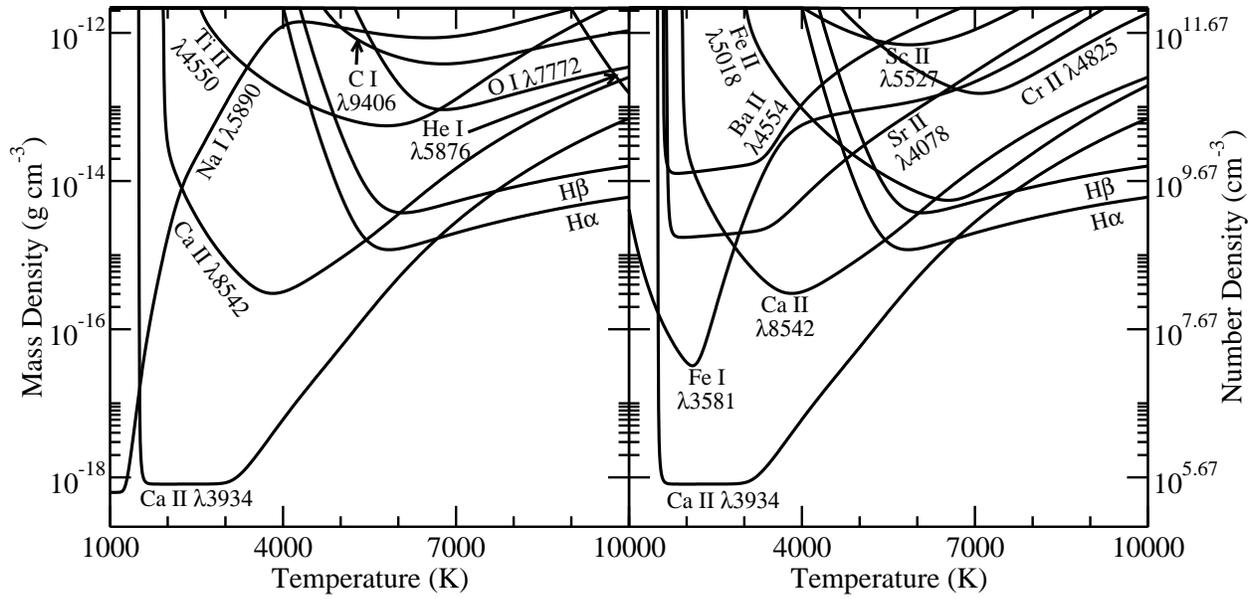}
\caption{
 Contours of $\tau = 1$ for a variety of ions in the H-rich composition
 \citep{Hatano1999b}, assuming TE.  The \caii, \halpha, and \hbeta\ 
 signature curves are repeated in both panels for comparison with 
 curves of other ions.
 \label{fig:refcont_h-rich}
}
\end{figure}

Figure \ref{fig:refcont_h-rich} is another ion signature plot, this
time using the H-rich composition \citep{Hatano1999b}.  The same
\caii\ signature curves are plotted, and the curves for both 
\halpha\ and \hbeta\ are included for comparison.
The required minimum density for an optical depth of 1 in \caii\ \wl
8542 is
45\% higher than in the C/O-rich case, and the density for which the
\halpha\ optical depth becomes unity is only 4 times higher than that.
This leaves a rather narrow density window in which to form the HV
\caii\ without also producing an \halpha\ signature, if the \halpha\
signature were to be produced under the assumption of TE.

It is an observational fact, however, that \halpha\ in SNe II is
generally in net or total emission \citep[see, for example, spectra of
SN 1999em --][]{Hamuy2001, Leonard2002, Elmhamdi2003}, 
an effect which cannot be
replicated by a direct analysis code which uses pure resonance
scattering line source functions where line flux is conserved except for
occultation effects.  However, pure resonance scattering is sufficient
for simultaneous, satisfactory fits of \hbeta\ and \hgamma\ in these
objects \citep[e.g.,][where synthetic \halpha\ is permitted to remain
too deep compared to observations in \Synow\ fits of SN 1999em]
{Baron2000}.

If the HV ejecta in SN 2000cx were a clump of HV H-rich material
blocking the photosphere, then the usual \halpha\ emission phenomenon
has an interesting consequence.  Assuming pure resonance scattering
for all the Balmer lines, this material would produce HV \halpha\
and \hbeta\ absorption features at the same velocities as the HV
\caii\ features.  The blueshifts are such that the HV \halpha\
absorption falls into the \ion{Si}{2} feature near 6100 \AA.  But the
net emission phenomenon will mitigate the strength of the \halpha\
absorption and further conceal the hydrogen signature.  We would
expect the \hbeta\ feature to remain in absorption, however.

\begin{figure}
\figurenum{13}
\epsscale{1.000}
\plotone{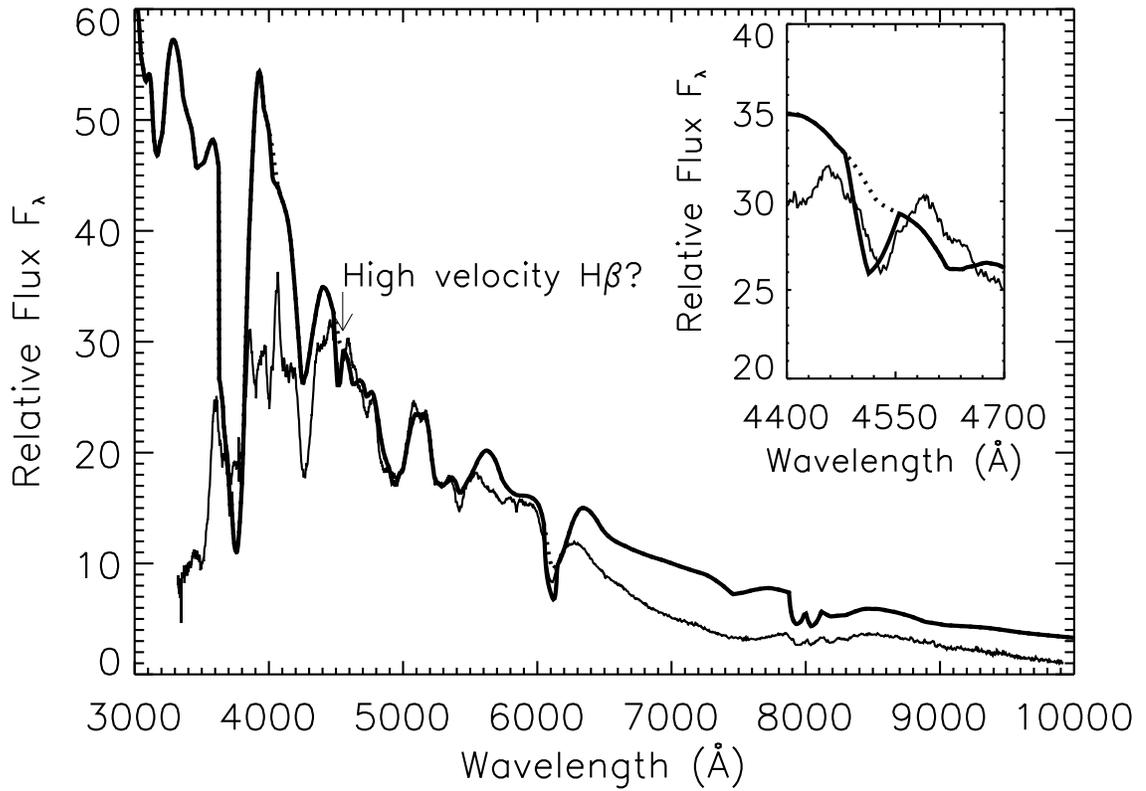}
\caption{
 \Brute\ fit to SN 2000cx spectrum at two days after maximum,
 with (solid line) and without (dotted line) H optical depth in the HV
 clump.  H$\alpha$ absorption is concealed by the Si II
 absorption at 6100 \AA.  The effect of H$\alpha$ could be completely
 erased by net emission effects.
 \label{fig:withHbeta}
}
\end{figure}

In Figure \ref{fig:withHbeta}, we show a fit to the entire spectrum
with and without hydrogen in the 3D HV ejecta.  The hydrogen reference line
(\halpha) optical depth used for this spectrum has a velocity-space
profile similar to that described in eq. (\ref{eq:cylRule})
except that the maximum optical depth at $v_z =22,400$ \kmps\ is
$\tau_{ref} = 4$.  As predicted, the assumption of pure resonance
scattering produces a readily apparent modification to the \ion{Si}{2}
absorption feature which would be ameliorated by the net emission
effect.  Additionally, a previously unidentified feature at 4500 \AA\
is fit by the synthetic HV \hbeta.

Note that this signature is only possible if the HV hydrogen is
confined to a clump blocking the photosphere.  If the HV hydrogen is
placed into a HV shell, a strong emission feature centered at the 
\halpha\ rest wavelength
will appear.  This is clearly not the case in the spectra of SN
2000cx.  Other candidate ions for fitting this line simply fail,
as discussed in a 1D direct analysis of this object (D. Branch et al., in
preparation).  This fit is included here since it is contingent
upon the 3D distribution of hydrogen at HV that we describe.

If the 4500 \AA\ feature is HV \hbeta, and we dispense with the assumption
of TE for the hydrogen lines, then we see that the lower level of
\hbeta\ could potentially be overpopulated.  This reduces the amount
of HV ejecta required to produce an \hbeta\ feature by a factor equal
to the departure coefficient of the lower level of this transition.
This reduces the gap between the HV masses estimated in this work and
those predicted by \citep{Marietta2000}.  But can NLTE effects bridge
this gap?  Future detailed 3D NLTE calculations are required to answer
this question.

Returning to Figure \ref{fig:refcont_h-rich}, we note the other ions
that could be detected if the HV ejecta were indeed H-rich.  It is
unfortunate that basically the same ions as in the C/O-rich case
appear.  The strongest discriminator between the two compositions is
the Balmer hydrogen series.  The majority of the hydrogen stripped
from the companion in the \citet{Marietta2000} models is at very low
velocity, so it seems that late-time spectra should reveal this
material.  But as \citet{Marietta2000} themselves point out, detection
of the material is problematic.  Nebular \halpha\ is severely blended
with iron and cobalt lines.  Resorting to the infrared introduces 
problems with atmospheric water lines (interfering with P$\alpha$), 
and other telluric and blending effects (interfering with P$\beta$).

However, \ion{He}{1} does make an appearance (in TE) at
high temperature.  If this ion is nonthermally excited, it might
produce an HV signature.  While \ion{He}{1} \wl 5876 is not observed,
nonthermal excitation could produce \ion{He}{1} \wl 10830.  At HV 
this line could explain a feature previously attributed by 
\citet{Rudy2001} to \ion{Mg}{2}.

Another scenario for the production of a SN Ia is through recurrent
novae \citep{Starrfield2003} or supersoft x-ray sources
\citep{Hachisu1999}.  These progenitors involve a He-rich (and
H-deficient) companion star.  Minimum mass estimates for this case
suggest HV ejecta masses about the same as in the H-rich case,
assuming TE.  But since helium stars are more tightly bound than
those simulated by \citet{Marietta2000}, mass stripping from a He-rich
companion seems less likely an origin for the HV material.

Arguments in favor of a circumstellar origin fall short.  The
spectroscopic signature for that material is completely different from
what is seen in the case of SN 2000cx.  In that case, we expect a
narrow H$\alpha$ emission spike, not unlike that found in SNe IIn
\citep[e.g.,][]{Filippenko1997}.


\section{Conclusions}

We have presented several exploratory fits to the unique \caii\ 
features of the unusual SN Ia 2000cx.  A 1D shell of material can
account for the HV IR triplet feature, but has difficulty doing so for
the corresponding \handk\ feature.  In 3D, both HV features can be fit
simultaneously with a chunk of material along the line of sight which
partially covers the photosphere.

Assuming a C/O-rich composition, mass estimates for the HV ejecta
in both geometries discussed are consistent in the lower limit with the
W7 model, suggesting a possible primordial origin.  If the HV ejecta
are H-rich, however, the mass required at HV to produce the observed
\caii\ features is somewhat higher.  There is circumstantial evidence
for the HV ejecta being H-rich, pointing to a signature of stripping
from a companion star.  This model is more favorable when NLTE effects
are taken into account, but confirmation requires detailed 3D spectrum
synthesis which is not currently feasible.

Understanding the origin of the HV ejecta in SNe Ia (and also its 
frequency and physical conditions) is potentially quite important for
future explosion models and for answering the progenitor question.  
It is clear that better constraints on the HV \ion{Ca}{2} phenomenon
require high-quality flux and polarization spectra
at near-infrared and near-ultraviolet wavelengths.

\acknowledgments

The authors acknowledge other members of the University of Oklahoma SN
Group, Dan Kasen and Peter Nugent, for their helpful comments.  The
research presented in this article made use of the SUSPECT\footnote{
http://www.nhn.ou.edu/$\sim$suspect} Online Supernova Spectrum Archive,
and the atomic line list of \citet{Kurucz1993}.  This
work has been supported by grant HST-AR-09544-01.A (provided by NASA
through the STScI, operated by the AURA, Inc., under NASA contract
NAS5-26555), NASA grant NAG5-12127, and NSF grants 
AST-9986965,
AST-9987438,
AST-0204771, and
AST-0307894.

\clearpage
\begin{deluxetable}{lccccc}
\tablecolumns{6}
\tablewidth{0pc}
\tablecaption{\Synow\ Fit Parameters for Non-Calcium Ions}
\tablehead
{
 Ion                     & 
 $\tau_{ref}( v_{min} )$ & 
 $v_{min}$               & 
 $v_{max}$               &
 $v_e$                   &
 $T_{exc}$               \\
                         & 
                         & 
 ($10^3$ \kmps)          & 
 ($10^3$ \kmps)          & 
 ($10^3$ \kmps)          & 
 ($10^3$ K)
}
\startdata
Fe III &  1.5 & 12.5 & $\infty$ & 1.0 & 10.0 \\
Si II  &  3.5 & 12.5 & $\infty$ & 1.0 & 10.0 \\
S II   &  1.5 & 12.5 & $\infty$ & 1.0 & 10.0 \\
O I    &  0.2 & 12.5 & $\infty$ & 3.0 &  8.0 \\
\enddata 
\label{tab:noCa}
\end{deluxetable}

\clearpage
\begin{deluxetable}{lcccccc}
\tablecolumns{7}
\tablewidth{0pc}
\tablecaption{\Synow\ \caii\ Fit Parameters}
\tablehead
{
 Fit                   & 
 (Figure)              & 
 Number of             & 
 $\tau_{ref}(v_{min})$ & 
 $v_{min}$             &
 $v_{max}$             &
 $v_e$                 \\
                       & 
                       & 
 Components            & 
	   & 
 ($10^3$ \kmps)        & 
 ($10^3$ \kmps)        & 
 ($10^3$ \kmps)
}
\startdata
1D1    & (\ref{fig:1d1})   & 1 &  1.4 & 12.5 & 30.0     & 20.0 \\ \hline
1D1PV  & (\ref{fig:1d1pv}) & 1 & 10.0 & 12.5 & $\infty$ &  3.0 \\ \hline
1D1HV  & (\ref{fig:1d1hv}) & 1 & 20.0 & 24.0 & $\infty$ &  0.5 \\ \hline
1D2    & (\ref{fig:1d2})   & 2 & 10.0 & 12.5 & 24.0     &  3.0 \\
       &                   &   & 30.0 & 24.0 & 25.0     &  3.0 \\ \hline
1D3    & (\ref{fig:1d3})   & 3 & 16.0 & 13.0 & 15.1     &  3.0 \\
       &                   &   &  7.0 & 19.0 & 23.5     &  3.0 \\
       &                   &   & 12.0 & 23.5 & $\infty$ &  3.0 \\
\enddata 
\label{tab:synow}
\end{deluxetable}

\clearpage

\clearpage

\clearpage

\clearpage

\clearpage

\clearpage

\clearpage

\clearpage

\clearpage

\clearpage

\clearpage

\clearpage

\clearpage

\end{document}